\documentclass[prd,twocolumn,superscriptaddress,preprintnumbers,nofootinbib,floatfix]{revtex4-2}
\usepackage{epsfig}
\usepackage{graphicx,physymb}
\usepackage{hyphenat}
\usepackage{amsmath}
\usepackage{comment}
\usepackage{amssymb}
\usepackage{mathtools}
\usepackage{mathrsfs}
\usepackage{slashed}
\usepackage{epstopdf}
\usepackage{xcolor}
\usepackage{braket}
\usepackage{booktabs}
\definecolor{lcolor}{rgb}{0.,0.0,0.}
\definecolor{citcolor}{rgb}{0,0.,0.5}
\usepackage[breaklinks,colorlinks,urlcolor=blue,citecolor=blue,linkcolor=blue]{hyperref}
\usepackage{multirow}
\usepackage{ltablex}
\usepackage{soul}
\usepackage[utf8]{inputenc}
\usepackage[T1]{fontenc}
\usepackage{latexsym}
\usepackage[normalem]{ulem}

\newcommand{\vb}[1]{\mathbf{#1}}

\newcommand{\pdv}[2][]{\frac{\partial{#1}}{\partial{#2}}}
\newcommand{\dd}[2][]{\mathrm d^{#1}{#2}\,} 
\newcommand{\dv}[2][]{\frac{\dd{#1}}{\dd{#2}}}
\newcommand{\Teps}{T_{\varepsilon}}

\newcommand{\beq}{\begin{eqnarray}}
\newcommand{\eeq}{\end{eqnarray}}

\def\dd{{\rm d}}

\newcommand{\bem}{\begin{multline}}
\newcommand{\eem}{\end{multline}}
\newcommand{\beg}{\begin{gather}}
\newcommand{\eeg}{\end{gather}}

\newcommand{\nn}{\nonumber\\}

\newcommand{\ben}{\begin{eqnarray*}}
\newcommand{\een}{\end{eqnarray*}}

\def\khat{\hat P} 
\def\jhat{\hat K} 
\def\rhat{\hat R}

\def\cP{{\cal P}}
\def\cK{{\cal K}}

\def\bs{\boldsymbol }

\def\bdel{\bs\partial}
\newcommand{\secn}[1]{Section~1}
\newcommand{\appn}[1]{Appendix~1}

\long\def\comment#1{ }

\def\and{\quad\text{and}\quad}

\newcommand{\rmd}{{\rm d}}

\newcommand{\rme}{{\rm e}}

\def\q{{\boldsymbol q}}
\def\0{{\boldsymbol 0}}

\def\k{{\vb k}}

\def\0{{\boldsymbol 0}}
\def\x{{\vb x}}
\def\y{{\vb y}}

\def\max{{\rm max}}

\newcommand{\qhat}{\hat{q}}

\newcommand{\tmin}{t_{\mathrm{min}}}
\newcommand{\tmax}{t_{\mathrm{max}}}

\begin{document}

\title{Jet quenching in out-of-equilibrium QCD matter  }

\author{Jo\~{a}o Barata}
\email{joao.lourenco.henriques.barata@cern.ch}
\affiliation{CERN, Theoretical Physics Department, CH-1211, Geneva 23, Switzerland}

\author{Kirill Boguslavski}
\email{kirill.boguslavski@subatech.in2p3.fr}
\affiliation{SUBATECH UMR 6457 (IMT Atlantique, Université de Nantes, IN2P3/CNRS), 4 rue Alfred Kastler, 44307 Nantes, France }
\affiliation{Institute for Theoretical Physics, TU Wien, Wiedner Hauptstraße 8-10, 1040 Vienna, Austria}

\author{Florian Lindenbauer}
\email{flindenb@mit.edu}
\affiliation{MIT Center for Theoretical Physics -- a Leinweber Institute, Massachusetts Institute of Technology, Cambridge, MA 02139, USA}
\affiliation{Institute for Theoretical Physics, TU Wien, Wiedner Hauptstraße 8-10, 1040 Vienna, Austria}

\author{Andrey V. Sadofyev}
\email{andrey.sadofyev@ehu.eus}
\affiliation{Laboratório de Instrumentação e Física Experimental de Partículas (LIP), Av. Prof. Gama Pinto, 2, 1649-003 Lisbon, Portugal}
\affiliation{Department of Physics, University of the Basque Country UPV/EHU, P.O. Box 644, 48080 Bilbao, Spain}
\affiliation{IKERBASQUE, Basque Foundation for Science, Plaza Euskadi 5, 48009 Bilbao, Spain}

\preprint{CERN-TH-2025-140} 
\preprint{MIT-CTP/5953}

\begin{abstract}
We present the first study of jet substructure modifications during the bottom-up evolution that describes the early stages of heavy-ion collisions. To this end, we study the \textit{bremsstrahlung} radiation rate of soft gluons from a hard parton propagating through out-of-equilibrium QCD matter. The gluon spectrum is computed within the Improved Opacity Expansion, which accounts for both multiple soft and single hard momentum exchanges between the hard probe and the medium. The background evolution is obtained from effective kinetic theory simulations that determine the jet quenching parameter, which in turn controls the radiation rate. We compute the radiation rate for initially under- and over-occupied systems, as well as for an expanding system undergoing hydrodynamization, which typically represents the initial stages of heavy-ion collisions. The results for these dynamical backgrounds are compared to static and thermally matched scenarios, allowing to gauge the importance of bulk expansion in the evolution of the jet cascade. Our findings show that the early stages of the bulk matter evolution in heavy-ion collisions leave a sizable imprint on the radiation pattern inside jets. These results establish a basis for incorporating pre-equilibrium dynamics into realistic descriptions of jet quenching and hard-probe evolution.
\end{abstract}

\maketitle

\section{Introduction}
High-energy heavy-ion collisions (HICs) are the primary experimental tool to probe QCD behavior under extreme conditions. The nuclear (bulk) matter produced in HIC experiments undergoes a multiphase evolution, beginning as a far-from-equilibrium state, before thermalizing into a nearly ideal liquid -- the quark-gluon plasma (QGP). After being formed, this hot and dense QCD medium expands and cools until the energy density drops below the hadronization transition, after which the resulting hadronic remnants are experimentally detected, see, e.g.~\cite{Busza:2018rrf, Berges:2020fwq, Apolinario:2022vzg} for reviews. These experiments offer a unique opportunity to study the out-of-equilibrium dynamics of QCD, thereby advancing our understanding of complex phenomena in quantum field theory.

Assuming a weak-coupling description, the currently favored picture of the medium evolution in HICs
is that the bulk matter produced immediately after  
the collision is initially dominated by highly occupied color fields, allowing for a description within classical Yang–Mills theory. In the context of the Color-Glass-Condensate (CGC) effective theory, this initial state of highly polarized color fields corresponds to the so-called \textit{Glasma}~\cite{McLerran:1993ni,Kovner:1995ts,Krasnitz:1999wc,Lappi:2006fp}, 
characterized by longitudinal chromo-electric and chromo-magnetic fields aligned along the beam axis, with a typical transverse size set by the saturation scale $Q_s$ inherited from the incoming nuclei. After a time of $\tau \sim 1/Q_s$, a quasiparticle picture emerges, and a kinetic theory description becomes applicable. The subsequent evolution drives the isotropization of the initially anisotropic system until it can be described within hydrodynamics. At weak coupling, these stages of equilibration are encompassed by the bottom-up thermalization scenario~\cite{Baier:2000sb,Berges:2013eia,Kurkela:2015qoa}, for further details on the (pre-)thermalization dynamics of QCD in the HIC context, we refer the interested reader to~\cite{Berges:2020fwq} for a review.

\begin{figure*}
    \centering
    \includegraphics[width=\linewidth]{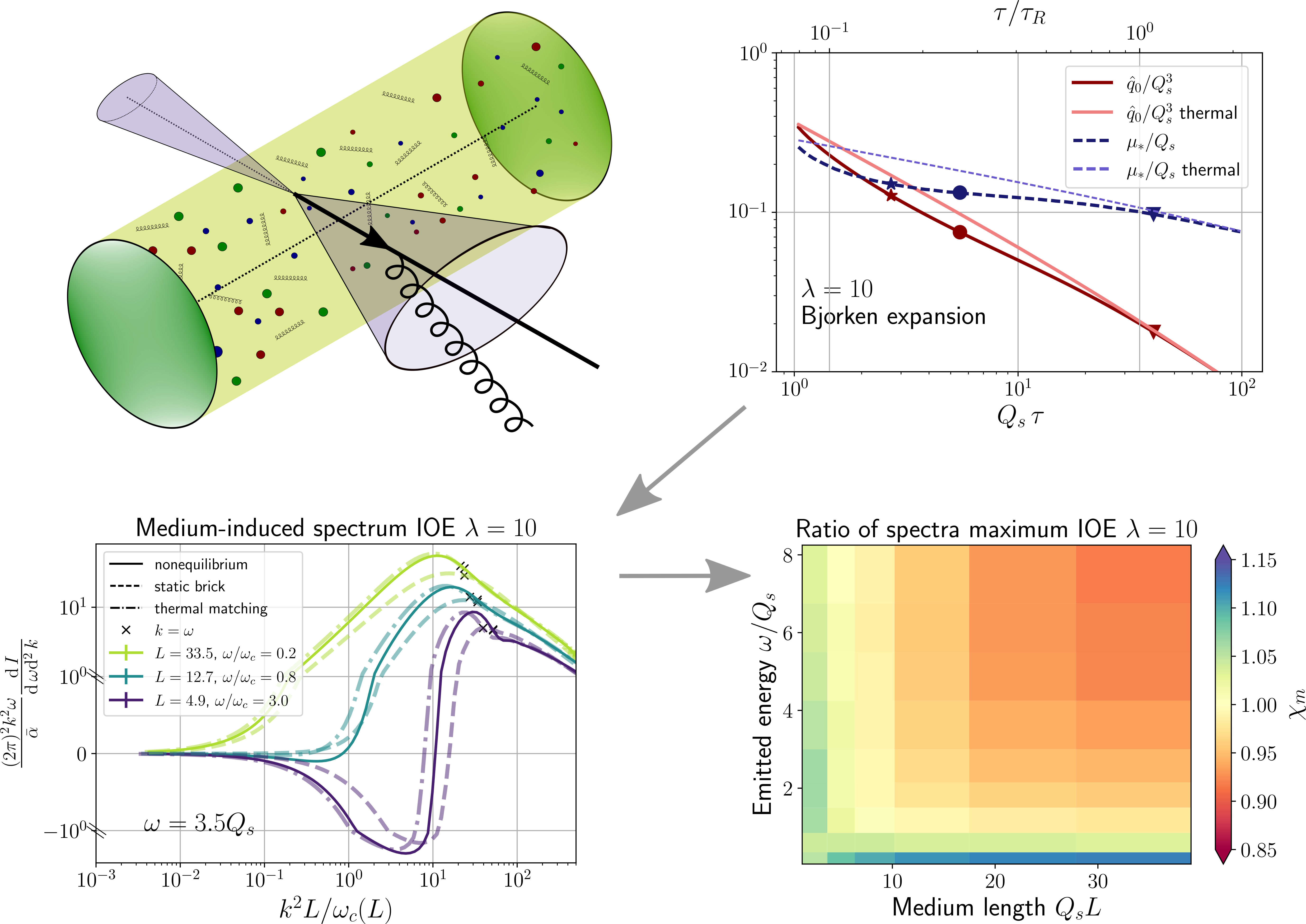}
    \caption{Summary of results for single gluon production in the presence of an out-of-equilibrium QCD medium, as illustrated on the top left panel. The bulk matter's time evolution (in green) is captured, for a Bjorken expanding system, by dynamical $\hat q_0$ and $\mu_*$, shown on the top right panel for a coupling $\lambda =10$. This allows to compute the gluon emission probability in the IOE as illustrated on the bottom left, and to compare to static and thermal matched results. These permit to gauge the relevance of the out-of-equilibrium bulk modifications during the jet evolution. In the remaining panel, we show the corresponding ratio of the spectra's maximum between the out-of-equilibrium and equilibrium scenarios: the lack of convergence to unity at late times points to the importance of the early stages in final state jet observables. 
    }
    \label{fig:summary_plot}
\end{figure*}

Despite the fact that this sophisticated theoretical picture for real-time thermalization dynamics in the aftermath of a HIC at weak coupling is broadly used, it remains unverified, as no experimental data can currently confirm or exclude it. This limitation arises because the early stages of HICs cannot be directly probed in the experiment, as these nonequilibrium states of matter never reach the detectors.
Instead, their properties must be inferred from the hadronic distributions of the QGP remnants, which are expected to retain imprints of the early-time evolution if not fully thermalized. Nonetheless, it is expected that such properties are not easily accessible through measurements of the soft final-state hadrons, since the late-time and long-lived hydrodynamic expansion tends to smear out this information. As a result, other probes of the early stages should be considered.
In this respect, it has been proposed that high-energy probes, such as QCD jets and heavy flavor, can be used to explore the early stages of HICs~\cite{Apolinario:2022vzg}. Since these energetic probes are produced and evolve in parallel with the bulk matter, they provide a powerful tool to construct a tomographic spacetime picture of the medium created in HICs. Moreover, their characteristic large energy or mass scale enables probing the bulk locally, while their spacetime evolution is not described within hydrodynamics. 

Putting the focus on jets, it has been argued that in the Glasma stage, the jet transport coefficient $\hat q$ ---
the key parameter governing medium-induced modifications to jets in most phenomenological studies --- can become substantially larger than its expected value in the QGP phase while also developing an anisotropic character~\cite{Ipp:2020mjc, Ipp:2020nfu, Avramescu:2023qvv,Carrington:2020sww,Carrington:2021dvw, Carrington:2022bnv,Avramescu:2025lhr, Backfried:2024rub}. Although the Glasma phase is short-lived, this raises the question whether early-stage jet quenching effects in HICs can become significant, see, e.g.,~\cite{Aurenche:2012qk,Adhya:2024nwx, Andres:2022bql, Singh:2025duj, Pablos:2025cli}.
Importantly, more recent studies focusing on the later pre-equilibrium dynamics described within effective kinetic theory (EKT)~\cite{Arnold:2002zm} report that $\hat q$ remains substantial during these intermediate stages~\cite{Boguslavski:2023alu, Boguslavski:2023waw, Boguslavski:2023jvg}, suggesting the possibility of sizable effects in jet observables~\cite{Barata:2024xwy}. Thus, to achieve a realistic description of jets in HICs, it is necessary to account for their interaction with the medium during the early stages of the evolution. This contrasts sharply with the conventional approach used in jet quenching phenomenology, where one typically assumes $\hat q = 0$ for timescales shorter than $\sim 1\, {\rm fm}/c$; see, e.g.,~\cite{Wang:2003mm}. Such assumptions should be revisited in detail to establish a solid framework for using jets as probes of the bulk matter in HICs, see~\cite{Pablos:2025cli,Adhya:2024nwx,Andres:2022bql} for recent related efforts. It is also worth emphasizing here that jet tomography of the early stages of matter in HICs may prove particularly insightful for the so-called small collisional systems, such as proton–nucleus or light–ion collisions. In these systems, the medium’s evolution is expected to be dominated by nonequilibrium dynamics and may not reach full hydrodynamization, thereby putting into question the applicability of conventional jet quenching frameworks that rely on near-equilibrium matter descriptions. Consequently, studying jets in the pre-equilibrium stage provides an essential link between the physics of hard probes in large and small hadronic systems.

In this work, we take a further essential step towards understanding how jets interact with nonequilibrium QCD matter by studying medium-induced gluon radiation in the pre-equilibrium phase. The modern description of the thermalization process  
in HICs is based on the EKT framework, which imposes a specific set of approximations from the outset~\cite{Arnold:2002zm}. Following this picture, we assume the medium to be homogeneous in the transverse plane, and describe it with a local EKT, which has been used in numerical studies of thermalization and hydrodynamization in HICs~\cite{Kurkela:2014tea, Kurkela:2015qoa}.

The in-medium branching process is analyzed in the dense matter limit, where soft gluon exchanges with the medium are resummed \cite{Baier:1996kr,Zakharov:1996fv,Baier:1996sk,Zakharov:1997uu}, while harder gluon exchanges are treated perturbatively \cite{Gyulassy:2000fs,Gyulassy:2000er}. We supplement that description with medium-induced color fields and their fluctuations, as dictated by the EKT of the bulk matter. We further neglect the collective flow or anisotropies in thermodynamic parameters on the jet quenching picture \cite{Sadofyev:2021ohn,Antiporda:2021hpk,Andres:2022ndd,Barata:2022krd,Barata:2022utc,Barata:2023qds,Barata:2023zqg,Kuzmin:2023hko,Barata:2024bqp,Kuzmin:2024smy,Bahder:2024jpa,Barata:2025htx} as these would take us beyond the common descriptions of the QCD matter thermalization, developing plasma instabilities in anisotropic matter \cite{Mrowczynski:1993qm, Romatschke:2003ms, Mrowczynski:2016etf}. Although such effects are expected to play a role in the physics of the early stages, 
a detailed analysis of these is left for future work. Under these assumptions, the medium-induced gluon spectrum becomes a functional solely of the bare jet quenching parameter $\hat q_0$ 
and an effective screening mass $\mu_*$.
We use the jet quenching parameter $\hat q$ extracted from EKT simulations~\cite{Boguslavski:2023waw, Boguslavski:2023alu, kurkela_2023_10409474}
to describe the properties of the medium.

We consider jet branching in under-occupied, over-occupied (both non-expanding), and longitudinally expanding systems. The first two cases represent simplified isotropic scenarios relevant to intermediate stages of the bottom-up thermalization picture~\cite{Baier:2000sb}, whereas the latter aims to capture the dynamics of HICs more realistically, where the jet evolves at mid-rapidity in an azimuthally symmetric bulk. The radiation spectrum is treated within the Improved Opacity Expansion (IOE) framework~\cite{Mehtar-Tani:2019tvy,Mehtar-Tani:2019ygg,Barata:2020rdn,Barata:2020sav,Barata:2021wuf,Adhya:2024nwx,Kuzmin:2025fyu}. This formalism can be straightforwardly combined with EKT simulations, while consistently accounting for both multiple soft and single hard interactions between the probe and the medium. This allows us to quantify how the medium evolution toward equilibrium impacts jet energy loss and substructure at leading order in the strong coupling, as illustrated in the top left panel of Fig.~\ref{fig:summary_plot}.

The main results of this work are summarized in the remaining three panels of Fig.~\ref{fig:summary_plot}: the top-right panel shows the time evolution of $\hat q_0$ and $\mu_*$ for a bulk undergoing Bjorken expansion, as obtained from EKT simulations. These parameters are then used to compute the medium-induced gluon spectrum (bottom left), which is directly compared to the results for the corresponding non-expanding static-brick and thermally fixed expanding backgrounds. This allows us to assess the relevance of the early out-of-equilibrium stages of HICs in the jet structure by, for example, comparing the highest peak of the gluon spectrum for the out-of-equilibrium and equilibrium scenarios, as illustrated in the bottom right panel. The observation that this ratio does not converge to unity at late times motivates one of the main conclusions of this paper: the early out-of-equilibrium stages of HICs can leave an imprint on final-state jet observables.

This work is organized as follows: In Section~\ref{sec:theory}, we introduce the theoretical foundations of our work, covering both the jet quenching and kinetic-theory aspects. In Section~\ref{sec:numerics}, we present numerical results on the computation of in-jet radiative processes in out-of-equilibrium matter. We conclude in Section~\ref{sec:conclusion} with a summary of our results, a discussion of their phenomenological implications, and an outlook on future research directions. 
Some more technical details of this work 
are presented in the appendix.

\section{Theoretical set up}\label{sec:theory}
In this section, we provide a brief overview of the soft medium-induced gluon spectrum originating from a hard parton in a dynamically evolving medium. Moreover, we discuss the jet quenching parameter obtained for out-of-equilibrium matter from EKT simulations, which is the relevant quantity to describe the bulk's dynamics.

\subsection{Bremsstrahlung gluon radiation}
We begin by detailing the soft-gluon emission process in a dense, homogeneous, and time-evolving QCD background. We aim to compute the probability of emitting a gluon with energy $\omega = z p_T$ and transverse momentum $|\mathbf{k}| \ll \omega$, off a hard source with energy $p_T \gg \omega$. In particular, we calculate the differential medium-induced gluon spectrum, which can be formally defined at leading order in the strong coupling constant as~\cite{Zakharov:1997uu}  
\begin{align}\label{eq:spectrum}
\frac{\rmd I}{\rmd \omega  \rmd^2 \k}&=\frac{\bar{\alpha}}{2\pi\omega^3} {\rm Re} \int_{t_0}^\infty \rmd t_2 \, \rme^{- \epsilon t_2} \int_{t_0}^{t_2}  \rmd t_1 \int_\x \,  \rme^{-i \k \cdot \x}\,  \nn
  & \hspace{-1 cm}\times \cP(\x,\infty;t_2) \, \bdel_\x\cdot \bdel_\y \cK(\x,t_2;\y,t_1)_{\y=0} - \frac{8\bar{\alpha}\pi}{\k^2\omega} \,, 
\end{align}
where $\bar{\alpha} = \alpha_s C_R / \pi$, with $C_R$ the Casimir factor of the parent parton taken to be a quark in this work ($C_R=C_F$), and $t_0$ controls the longitudinal position (time) at which it is produced. 
Here we use the notation $\int_\x = \int \rmd^{2}\x$ and $\int_\k = \int {\rmd}^2 \k (2\pi)^{-2}$ for transverse integrations, and we introduce an adiabatic turn-off $\epsilon$ to remove interactions at asymptotic distances~\cite{Wiedemann:2000za}.

The soft-gluon 
spectrum is fully determined by the functions $\cP$ and $\cK$, which encapsulate all the relevant medium parameters. The former describes the momentum broadening of the final state gluon, while the latter characterizes its emission  
from the quark. Under the common assumptions employed in jet quenching, they can be expressed in terms of 
specific correlation functions of light-like Wilson lines,  
see e.g.~\cite{Mehtar-Tani:2013pia, Casalderrey-Solana:2007knd, Isaksen:2023nlr} for a detailed discussion.
Assuming that the background field describing the medium follows Gaussian statistics, with correlations that are local in space and time and diagonal in color, and that the medium is homogeneous in the transverse directions, these objects satisfy
\begin{align}\label{eq:def_P_and_K}
    &{\hspace{1. cm}}\partial_t \cP(\x,t;t_1) = -  v(\x,t)\cP(\x,t;t_1) \, ,\nn
    &{\hspace{0 cm}}\left(i\partial_t-\frac{\partial^2_\x}{2\omega} + i v(\x)\right) \cK(\x,t;\y,t_1)\notag\\
    &\hspace{2.8cm}= i \delta^{(2) }(\x-\y) \delta(t_1-t) \, .
\end{align}
Here, $v(\x,t)$ is an effective scattering 
potential that incorporates the microscopic details of the interactions between the high-energy partons and the soft modes of the medium. 
At leading logarithmic accuracy, it is known to have a universal form
\begin{align}
    v(\x,t) = \frac{\hat q_0(t)}{4} \x^2 \log \frac{1}{\x^2 \mu_\ast^2(t)} +\mathcal{O}(\hat q_0 \x^4 \mu_\ast^2 ) \, , 
    \label{eq:potential_expansion}
\end{align}
where we will refer to $\hat q_0$ as the bare jet quenching parameter, and $\mu_\ast$ is the characteristic screening mass. Both of these parameters are model-dependent when written in terms of the medium scales, but as shown below, the structure of the final results does not depend on these details.
We shall discuss in the next subsection how they can be extracted from a QCD kinetic theory simulation.
It will also be useful to introduce the (full) jet quenching parameter at a scale $Q$ as
\begin{align}
    \hat q(t) \equiv \hat q_0(t)\log\frac{Q^2}{\mu_\ast^2(t)}\,,
\end{align}
which combines both quantities.

An analytic closed-form solution for $\cP$ can be readily obtained for a generic potential. This is, however, not the case for $\cK$, whose closed-form solution is only known for specific choices of $v(\x,t)$. Analytical approaches to derive $\cK$ usually either focus on the dilute regime, see \cite{Sievert:2018imd} and references therein, expanding in powers of $v(\x,t)$ (so-called opacity expansion), or consider the resummed form under the harmonic oscillator (HO) approximation for $v(\x,t)$, see \cite{Casalderrey-Solana:2007knd}, accounting only for the leading contributions in Eq.~\eqref{eq:potential_expansion}.  
Although one can attempt to solve for this kernel in general, the range of the solutions is limited by the inherent complexity of Eq.~\eqref{eq:def_P_and_K}, see e.g.~\cite{Caron-Huot:2010qjx,Isaksen:2023nlr}.

In this work, we circumvent these issues by 
employing the IOE, which allows us to treat the gluon spectrum analytically. By expanding around the HO limit, corresponding to a quadratic form of $v(\x)$, and treating the $\x$-dependent part of the logarithm in Eq.~\eqref{eq:potential_expansion} as a perturbation, we obtain analytic expressions for the radiative spectrum that account for both the regime dominated by multiple soft-gluon exchanges and the large-momentum sector in which single hard interactions with the background govern the dynamics.
Not only is the approach followed in the IOE computationally and theoretically useful, but also the all-order structure of the expansion has been studied in detail~\cite{Barata:2020sav}. In particular, it has been argued that the relevant qualitative features of the exact spectrum are already captured by the leading and next-to-leading order terms of the IOE.%
\footnote{This observation was originally made for the energy spectrum. However, it is generally expected to extend to other observables as well.}

To apply the IOE at the level of the radiation rate, one first rewrites the scattering potential at leading logarithmic accuracy, see Eq.~\eqref{eq:potential_expansion}, as
\begin{align}
  v(\x) &=   \frac{\hat q_0(t)}{4} \x^2 \log \frac{Q^2}{\mu_\ast^2} - \frac{\hat q_0(t)}{4} \x^2 \log  Q^2 \x^2  \nn 
  &= v_{\rm HO}(\x,t) + \delta v(\x,t)\, ,
\end{align}
for some $Q^2\gg \mu_\ast^2$ and with the identification $v_{\rm HO}(\x,t) = \frac{1}{4} \x^2 \hat q(t)$. In this framework, neglecting $\delta v$, yields the HO approximation. 
The leading-power IOE then corresponds to expanding $\cP$ and $\cK$ separately up to $\mathcal{O}(\delta v)$. 
As
shown in 
detail in~\cite{Barata:2020sav}, the term in the expansion that depends only on $v_{\rm HO}$ describes the physics of multiple soft scatterings in the medium, while hard Coulomb interactions are encoded in $\delta v$. Since $\delta v$ depends on the transverse coordinate $\x$, one has to ensure that the expansion remains valid for each specific observable, as the integration over $\x$ can, in principle, probe regions where the correction is not negligible. Thus, the IOE 
becomes consistent only after imposing a constraint on the matching scales, as detailed below. By applying this expansion in Eq.~\eqref{eq:spectrum}, the radiation spectrum can be written as 
\begin{align}
   \frac{\rmd I}{\rmd \omega  \rmd^2 \k} =\frac{\rmd I^{(0)}}{\rmd \omega  \rmd^2 \k} + \frac{\rmd I^{(1)}}{\rmd \omega  \rmd^2 \k} \, ,\label{eq:spectrum-splitting-lo-nlo}
\end{align}
where the leading-order term 
corresponds to the solution 
obtained with the harmonic potential $v_{\rm HO}$:
\begin{align}\label{eq:lo-spectrum}
     \frac{\dd I^{(0)}}{\dd \omega \dd ^2\k} &=  \frac{2 \bar \alpha }{\pi\omega} {\rm Re} \, \Bigg\{  \k^{-2}\left( e^{-i \frac{\k^2}{2\omega   {\rm Cot}(L,t_0)}}-1\right) \nn 
    &+   \int_{t_0}^L \dd t\, \frac{  {\rm Cot}(t,t_0)}{\hat P^2(t,t_0)} e^{-\frac{\k^2}{\hat P^2(t,t_0)}} \Bigg\}\, ,
\end{align}
and $L$ denotes the total longitudinal distance traversed by the leading parton in the medium. The time-dependent functions entering the spectrum can be constructed from the solutions of the corresponding initial-value problem
\begin{align}
&\left[\frac{\rmd^2}{\rmd t^2}+\Omega^2(t)\right]F(t,t_0)=0 \, ,\label{eq:differential-equation}
\end{align}
where $F=\{S,C\}$ with the initial conditions $S(t_0,t_0)=\partial_t  C(t,t_0)_{t=t_0} =0$, $\partial_t  S(t,t_0)_{t=t_0}=C(t_0,t_0)=1$. 
The effective 
frequency is defined as 
\begin{align}
\label{eq:Omega_qhatr}
\Omega(t)=\frac{1-i}{2}\sqrt{\frac{\hat{q}_r(t)}{\omega}},
\end{align}
where $\hat q_r$ denotes the full jet quenching parameter at an associated soft-gluon radiation scale $Q_r$, which we specify below. 
It is also convenient to introduce further shorthand notations:
\begin{align}
&{\rm Cot}(t_2,t_1)\equiv \frac{C(t_1,t_2)} {S(t_2,t_1)}=\frac{C_{12}}{S_{21}} = {\rm Cot}_{21}\,, \nn
&\khat^2_{21} = Q^2_b(L,t_2) -2i\omega{\rm Cot}_{21} \, ,\nn
&\jhat_{21} =\frac{i}{2\omega}\left(-{\rm Cot}_{12}+{\rm Cot}_{10}+\frac{2i\omega}{S^2_{21}\khat^2_{21}}  \right)S_{21}^2 \khat_{21}^4\, , \nn 
&\rhat_{21} =-\frac{2i\omega}{\khat^2_{21}S_{21}}\, .
\end{align}

We now define three characteristic momentum scales that appear in the IOE construction. The first one, 
\begin{align}
Q_{s0}^2(t_2,t_1)=\int_{t_1}^{t_2} \rmd t \, \hat{q}_0(t), 
\end{align}
represents the accumulated transverse-momentum broadening from the bare quenching parameter. The second scale, governing diffusive transverse-momentum broadening and associated with $\cP$, is defined as
\begin{align}\label{eq:Qb_final}
Q_b^2=\displaystyle\int_0^{L} \rmd t \,  \hat q_0(t)\log\frac{Q^2_b}{\mu_\ast^2(t)} \equiv \displaystyle\int_0^{L} \rmd t \,\hat q_b(t) \, .
\end{align}
Similarly, an analogous scale for $\cK$ can be introduced as 
\begin{align}\label{eq:Qr_final}
Q^2_r(t)=\sqrt{\hat q_0(t)\,\omega \log\frac{Q^2_r}{\mu^2_\ast(t)}} \equiv \sqrt{\omega\,\hat q_r(t)} \, ,
\end{align}
which 
controls the production of soft gluons in the medium. These scales serve as self-consistency conditions of the IOE, ensuring that both $\cP$ and $\cK$ can be treated perturbatively in $\delta v$, {\it i.e.}, that the expansion around the harmonic potential remains valid. With this definition, 
one has $\hat q_r = \hat q_0 \log (Q_r^{2} \mu_\ast^{-2})$, 
and $\hat q_r$ and $\hat q_b$ 
represent the jet quenching parameter evaluated at different matching scales. 
For the radiation spectrum derived under the IOE, it is natural to identify $\hat q_r$ as the relevant jet quenching parameter governing medium-induced gluon-emission. 
Once the IOE prescription is specified, the definitions of the matching scales are uniquely fixed for each observable by their respective self-consistency conditions in Eqs.~\eqref{eq:Qb_final} and \eqref{eq:Qr_final}. When real solutions to the matching scale $Q_r$ cannot be found, the IOE ceases to be applicable, as expected at very low gluon energies, close to the characteristic Bethe-Heitler frequency, $\omega \sim \mu_*^4/\qhat_0$, where the multiple-scattering approximation underlying the expansion breaks down~\cite{Barata:2021wuf}.

The next-order terms in the IOE were computed for a generic time-dependent medium in~\cite{Barata:2021wuf}, where explicit expressions were derived for arbitrary $\hat{q}$ profiles.
For practical applications---since $\hat q_0$ must vanish at late times---it is convenient to consider a system in which $\hat{q}_0$ evolves non-trivially within a finite time interval, while outside this region it is set to zero $\hat{q}_0=0$.
In this case, the general expressions of~\cite{Barata:2021wuf} can be further simplified analytically, yielding the compact form: 
\begin{widetext}
\begin{align}
\frac{\rmd I^{(1)}}{\rmd \omega  \rmd^2 \k} &=\frac{\bar{\alpha}\pi}{\k^4}   {\rm Re}  \int_{t_0}^L \rmd t_2 \Bigg\{ \frac{i}{2\omega}\, \,\int_{t_0}^{t_2} \rmd t_1 \,  \frac{\hat{q}_0(t_1)}{\rhat_{21}^2} \rme^{-\frac{\k^2}{\khat^2_{21}}} 
  \left[Q^2_s(L,t_2) I_a\left(\frac{\k^2}{\jhat_{21}},\frac{\k^2\rhat^2_{21}}{Q_r^2}\right)+2 C_{12}\rhat_{21} \k^2 I_b\left(\frac{\k^2}{\jhat_{21}},\frac{\k^2\rhat^2_{21}}{Q_r^2}\right)\right] \nn 
  & -  {\rm Cot}_{20}\, Q^2_{s0}(L,t_2) \, I_a\left(\frac{\k^2}{\khat^2(t_2,t_0)},\frac{\k^2}{Q_b^2}\right)\nn 
   &+2   \,  \hat q_0(t_2) C^2(t_2, L) I_b\left(\frac{i\k^2}{2\omega C^2(t_2 ,L)\left({\rm Cot}_{20}-\frac{1}{C(t_2 ,L)S(L,t_2)}-{\rm Cot}(t_2, L)\right)}; \frac{\k^2}{Q_r^2  C^2(t_2,L)}\right) e^{-\frac{i\k^2}{2\omega {\rm Cot}(L,t_2)}}\Bigg\} \, ,\label{eq:nlo-spectrum}
\end{align}
\end{widetext}
where we have introduced the shorthand notations
\begin{align}
I_a(x,y ) &=  \int_{0}^\infty \rmd z \, z^3 \,\log\frac{y}{z^2}   \, J_0(z) \,\rme^{ - \frac{z^2}{4x}}\, , \nn  
I_b(x,y )&=\int_{0}^\infty \rmd z \, z^2 \log \frac{y}{z^2} J_1\left(z \right)\rme^{-\frac{z^2}{4x}} \, ,
\end{align}
which admit simple analytic representations, see~\cite{Barata:2021wuf, Kuzmin:2025fyu}.

\subsection{Out-of-equilibrium QCD matter}
As discussed in the previous section, to evaluate the gluon radiation spectrum, one needs to specify $\hat q_{r/b}(t)$ for the leading term in the IOE, 
and $\hat q_0(t)$ and $\mu_\ast(t)$ to describe the complete series. Here, we outline how these parameters can be obtained from the EKT. We closely follow the discussion in~\cite{Boguslavski:2023waw}.

The effective kinetic description of QCD is valid at weak couplings or sufficiently high energies and assumes that the bulk medium admits a quasiparticle description \cite{Arnold:2002zm}. This framework is also commonly used to extrapolate to phenomenological couplings relevant in the HIC context. In kinetic theory, all information about the system is encoded in the particle distribution function $f(\vec p,t)$, whose time evolution is governed by 
the Boltzmann equation with elastic and inelastic collision kernels, 
\begin{align}
\pdv[f(\vec p,\tau)]{\tau}- \frac{p_z}{\tau} \pdv[f(\vec p,\tau)]{p_z} =-\sum_{i=1,2} \mathcal{C}^{2\leftrightarrow i}[f(\vec p,\tau)] \, ,\label{eq:boltzmann_equation}
\end{align}
where we denote three-dimensional vectors with an arrow, i.e., $\vec p$ to visually distinguish them from two-dimensional vectors, e.g., $\k$.
Here, the second term accounts for the expansion  
in HICs \cite{Mueller:1999pi} and is absent in non-expanding setups, which we also study in this work. Correspondingly, we will use the symbol $\tau$ to label the (proper-)time for the expanding systems, while using the time $t$ for the isotropic cases.
For simplicity, we will consider a purely gluonic plasma,%
\footnote{Accounting for quarks in the kinetic description is straightforward and can be found in the literature, see e.g.~\cite{Kurkela:2018xxd, Du:2020zqg}.}
since during the early stages of HICs, when hydrodynamization occurs, gluons are the dominant degrees of freedom.

Starting from a potentially far-from-equilibrium initial distribution, Eq.~\eqref{eq:boltzmann_equation} can be solved numerically to study the system’s out-of-equilibrium evolution. Building on this framework, QCD kinetic theory simulations have been employed to investigate equilibration in isotropic~\cite{Kurkela:2014tea, Fu:2021jhl} and expanding systems~\cite{Kurkela:2015qoa, Kurkela:2018xxd, Du:2020zqg}. This framework has been broadly used in phenomenological studies, see e.g.~\cite{Kurkela:2018wud,
Garcia-Montero:2023lrd, Garcia-Montero:2024lbl, Zhou:2024ysb, Boguslavski:2025ylx}.

Here, we focus on three representative setups for the evolution of the bulk:
\begin{enumerate}
    \item \textbf{Isotropic, initially under-occupied system:} We start by studying a system whose gluon density is initially smaller than in equilibrium. Following \cite{Kurkela:2014tea}, we use a Gaussian distribution centered at $p=Q$ as the initial condition,
\begin{align}
    f(p,t_0)=A\exp\left(-100\, (p-Q)^2/Q^2\right)\, ,
\end{align}
where the proportionality constant $A$ is 
fixed such that $Q=50T$, with $T$ being 
the temperature of the equilibrium system after thermalization.
\item \textbf{Isotropic, initially over-occupied system:} 
Here, we consider a plasma with an initially large gluon density.
As the initial condition, we employ the scaling solution that corresponds to the non-thermal far-from-equilibrium attractor, where the distribution function exhibits a self-similar scaling form \cite{Kurkela:2012hp, Schlichting:2012es, Berges:2013fga, Arnold:2005ef, Berges:2008mr}. 
This form can be
parametrized as~\cite{AbraaoYork:2014hbk}
\begin{align}
    f(p,t_0)&=\frac{(Qt_0)^{-4/7}}{\lambda \tilde p}\nn 
    &\times \left(0.22 e^{-13.3\tilde p}+2 e^{-0.92 \tilde p^2}\right),
\end{align}
with $\tilde p=(p/Q)(Qt_0)^{-1/7}$ and the initialization time $Qt_0=10^{-4}$.
\item \textbf{Expanding system:} 
We further focus on the system considered in~\cite{Boguslavski:2023alu, lindenbauer_2023_10419537}, where $\hat q(\tau)$ has been extracted during the bottom-up thermalization process, as described below. There, the evolution was initialized with a distribution inspired by the CGC effective theory~\cite{Kurkela:2015qoa}, which captures the early-time momentum anisotropy of the Glasma via
\begin{align}  
    f(p_\perp,p_z,\tau{=}1/Q_s)=\frac{2A(\xi)\langle p_T\rangle}{\lambda \, p_\xi}e^{-\frac{2p_\xi^2}{3\langle p_T\rangle^2}},
    \label{eq:initial_cond_expanding}
\end{align}
and whose parameters had been obtained from a JIMWLK-evolved 2+1D Glasma simulation~\cite{Lappi:2011ju}. The resulting two-dimensional distribution is then broadened using an anisotropy parameter $\xi$ via $p_\xi=\sqrt{p_\perp^2+(\xi p_z)^2}$, where we choose
$\xi = 10$, $A(\xi
) =  5.24171$, and $\langle p_T\rangle = 1.8\, Q_s$, in accordance with previous studies.

\end{enumerate}

For the isotropic kinetic theory simulations, i.e., the initially under- and over-occupied conditions described above, we use the hard-thermal-loop screened matrix elements (isoHTL screening) as described in~\cite{Boguslavski:2024kbd}. For the case of an expanding system, we use the results of~\cite{Boguslavski:2023alu, lindenbauer_2023_10419537}, which were obtained from QCD kinetic theory simulations employing the Debye-like screening prescription~\cite{AbraaoYork:2014hbk, Boguslavski:2024kbd}. In all of these cases, the jet quenching parameter $\hat q(t)$ is obtained from an integral over the distribution function, see e.g. \cite{Boguslavski:2023alu, Boguslavski:2023waw},
\begin{align}
\hat q &= \lim_{p\to\infty}\int_{q_\perp < \Lambda_\perp}\!\!\!\!\!\!\!\!\!\dd\Gamma\, q_\perp^2 (2\pi)^4\delta^{(4)}(P+K-P'-K') \nonumber \\
&\qquad\times \frac{1}{2 \nu}\,\frac{\left|\mathcal M\right|^2}{p} f(\vec k)\left(1+ f(\vec k')\right),
\label{eq:qhat_general}
\end{align}
where
$|\mathcal M|^2$ 
denotes the pure gluonic scattering matrix element with the internal soft-gluon line replaced by the isotropic hard-thermal-loop propagator~\cite{Boguslavski:2023waw}, and the integral measure is defined as 
\begin{align}
\int_{q_\perp < \Lambda_\perp}\!\!\!\!\!\!\!\!\!\dd\Gamma=\int\frac{\dd^3\vec p}{(2\pi)^32|\vec p|}\frac{\dd^3\vec k}{(2\pi)^32|\vec k|}\frac{\dd^3\vec k'}{(2\pi)^32|\vec k'|}\Theta(\Lambda_\perp -q_\perp),
\end{align}
with $\vec q=\vec p'-\vec p=\vec k-\vec k'$. Here, $\Theta(x)$ is the usual step function, which enforces a transverse momentum cutoff $q_\perp<\Lambda_\perp$ in the integral. The uppercase letters denote the light-like four-momenta of the particles considered in the scattering process, e.g., $K^\mu=(|\vec k|,\vec k)$. The quantity $\nu=2(N_c^2-1)$ counts the number of gluonic degrees of freedom.
It can be shown that Eq.~\eqref{eq:qhat_general} exhibits a logarithmic dependence on the transverse-momentum cutoff $\Lambda_\perp$ \cite{Boguslavski:2023waw}. 
Following~\cite{Boguslavski:2023alu}, we express this dependence in the form
\begin{align}
    \hat q(\Lambda_\perp\gg T_\varepsilon,t)=a(t)\log\frac{\Lambda_\perp}{Q_s}+b(t)\, , \label{eq:qhat_parametrization}
\end{align}
where $Q_s$ sets the momentum scale relative to which $\Lambda_\perp$ is measured (in \cite{Boguslavski:2023alu} this corresponds to the saturation momentum). We now aim to use this parametrization to determine $\hat q_0$ and $\mu_\ast$ in Eq.~\eqref{eq:potential_expansion}, which are required as input for calculating the radiation spectrum. 

The relation between $\hat{q}$ and the small-distance behavior of the potential has been broadly discussed in the literature, see e.g.~\cite{Baier:1996sk,Arnold:2008vd,Arnold:2008iy}, but has only recently been described in terms of the logarithmic cutoff-dependence of Eq.~\eqref{eq:qhat_parametrization} \cite{Altenburger:2025iqa}. Following this logic, we start from $v(\x)$, which determines $\hat q$ and whose short-distance expansion allows one to identify $\hat q_0$ and $\mu_\ast$ as in Eq.~\eqref{eq:potential_expansion}. The potential can be expressed as
\begin{align}
	v(\x)=\int\frac{\dd^2{\vb q}}{(2\pi)^2}(1- e^{i\x\cdot\vb q})C(\vb q)\,,\label{eq:fourier}
\end{align}
where $C(\vb q)$ is the collision kernel (or momentum-broadening probability). 
The jet quenching parameter then follows as its second moment
\begin{align}
    \hat q=\int\frac{\dd^2{\vb q}}{(2\pi)^2}{\vb q}^2 C(\vb q)\,.
\end{align}
To analyze the small-$\x$ limit of $v(\x)$, it is convenient to separate the integration region into soft ($q <\Lambda_\perp$) and hard ($q>\Lambda_\perp$) parts.
The potential can then be rewritten as
\begin{align}
	v(\x)&=\int_0^{\Lambda_\perp} \frac{\dd^2{\vb q}}{(2\pi)^2}(1-e^{i\x \cdot\vb q})C(\vb q)\nn
    &+\int_{\Lambda_\perp}^\infty \frac{\dd{q}}{2\pi} \, q \, C(q)(1-J_0(xq)),
\end{align}
where for large $q=|\q|$ we can use the isotropic perturbative form
\begin{align}
C(q\gg T)=\frac{2 C_A^2g^4 n}{\nu}\frac{1}{q^4},
\end{align}
which allowed us to perform the angular integral analytically.
Here, $n=\nu\int\frac{\dd^3{\vec p}}{(2\pi)^3}f(\vec p)$ is the particle number density. Expanding both integrals in  $\x$, one finds
\begin{align}
    \int_0^{\Lambda_\perp} \frac{\dd^{2}{\vb q}}{(2\pi)^2}(1-e^{i\x\cdot\vb q})C(\vb q)=\frac{1}{4}\x^2\hat q({\Lambda_\perp}) + \mathcal O(\x^4)\,,
\end{align}
and
\begin{align}
    &\int_{\Lambda_\perp}^\infty \frac{\dd{q}}{2\pi}q\, C(q)(1-J_0(xq))\notag \\
    &= \x^2\frac{C_A^2 g^4 n}{4\pi\nu}\left(1-\gamma_E-\log\frac{x {\Lambda_\perp}}{2}\right) + \mathcal O\left(\x^4\right)\,.
\end{align}
Requiring that the small-$\x$ expansion of Eq.~\eqref{eq:fourier} be independent of the cutoff ${\Lambda_\perp}$ then yields
\begin{align}
	a(t)=\frac{C_A^2 g^4 n(t)}{\pi\nu}\label{eq:universalform_logcoefficient},
\end{align}
which is precisely the form obtained in~\cite{Boguslavski:2023waw} for isotropic distributions. Comparing to Eq.~\eqref{eq:potential_expansion}, we notice that $\hat q_0=a/2$, then
\begin{align}
    v(\x)=\frac{1}{4}\hat q_0
    \x^2\left(\frac{b}{\hat q_0}+2-2\gamma_E+\log\frac{4}{\x^2Q_s^2}\right) \, .
\end{align}
Note that, unlike the jet quenching parameter $\hat q({\Lambda_\perp})$, this expression is independent of any cutoff in the integral. Moreover, this derivation clarifies how the cutoff dependence of $\hat q({\Lambda_\perp})$ should be used to determine the small-distance limit of the potential, which provides the physical input entering Eq.~\eqref{eq:def_P_and_K}.
It was also shown in~\cite{Altenburger:2025iqa} that this expression numerically reproduces the short-distance behavior of $v(\x)$, and from Eq.~\eqref{eq:potential_expansion} we readily obtain 
\begin{align}
	\mu_\ast ^2=\frac{1}{4}e^{-\frac{b}{\hat q_0}+2\gamma_E-2} \, .
\end{align}

\begin{figure*}
    \centerline{
        \includegraphics[width=0.5\linewidth]{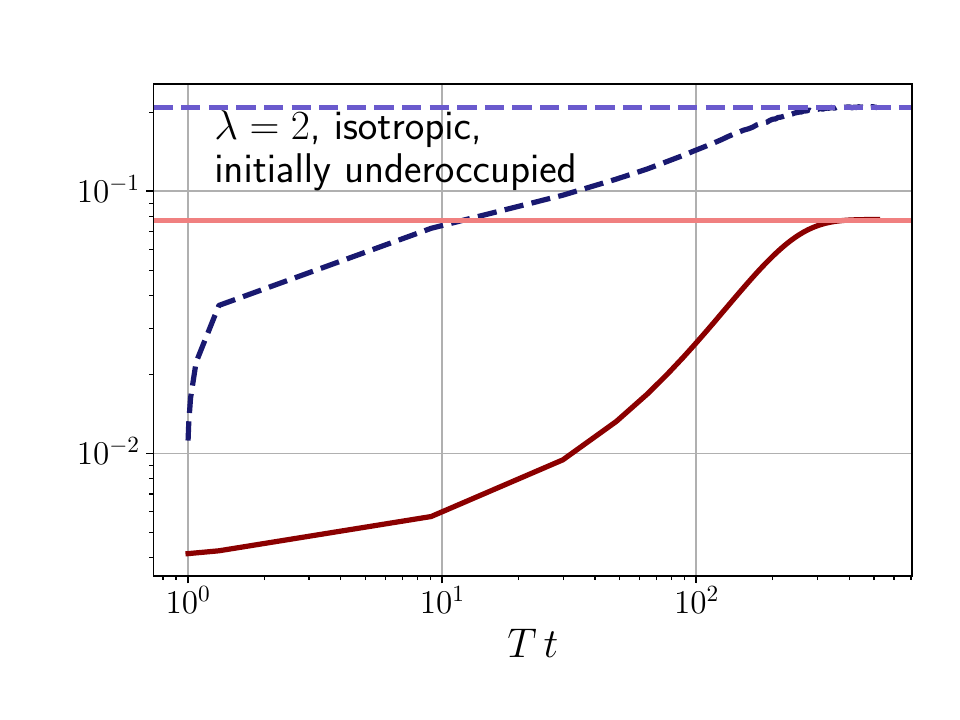}
        \includegraphics[width=0.5\linewidth]{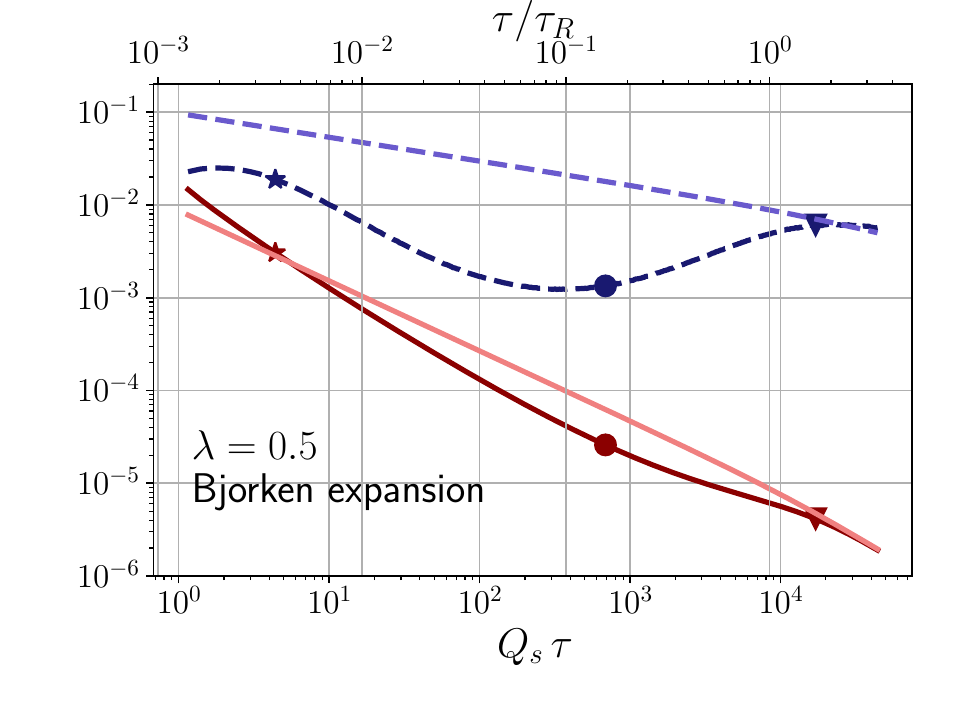}
    }
    \centerline{
        \includegraphics[width=0.5\linewidth]{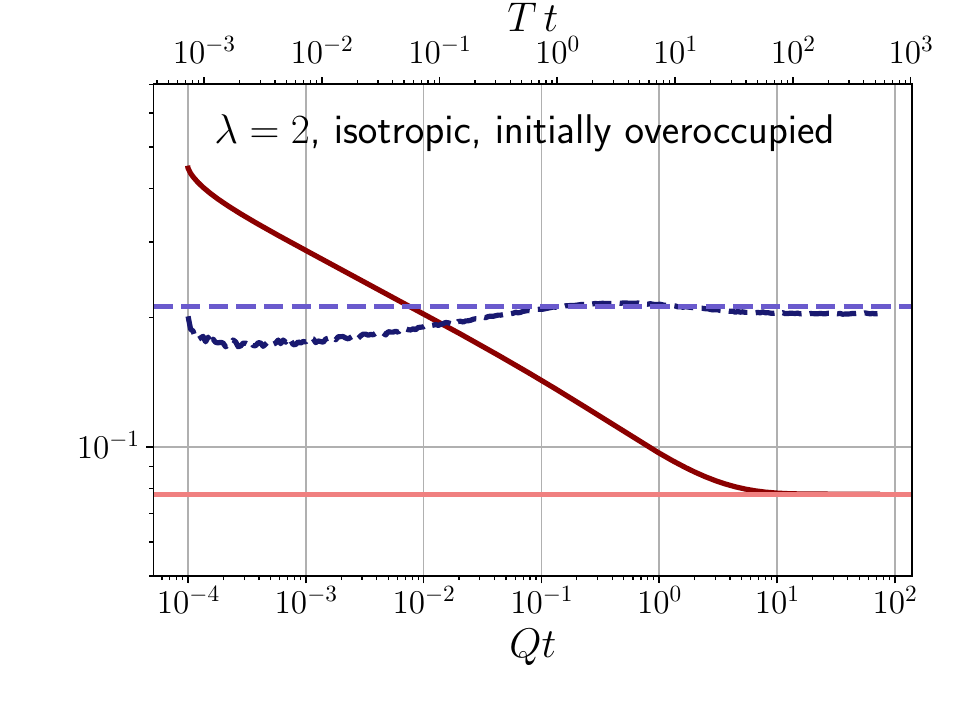}
        \includegraphics[width=0.5\linewidth]{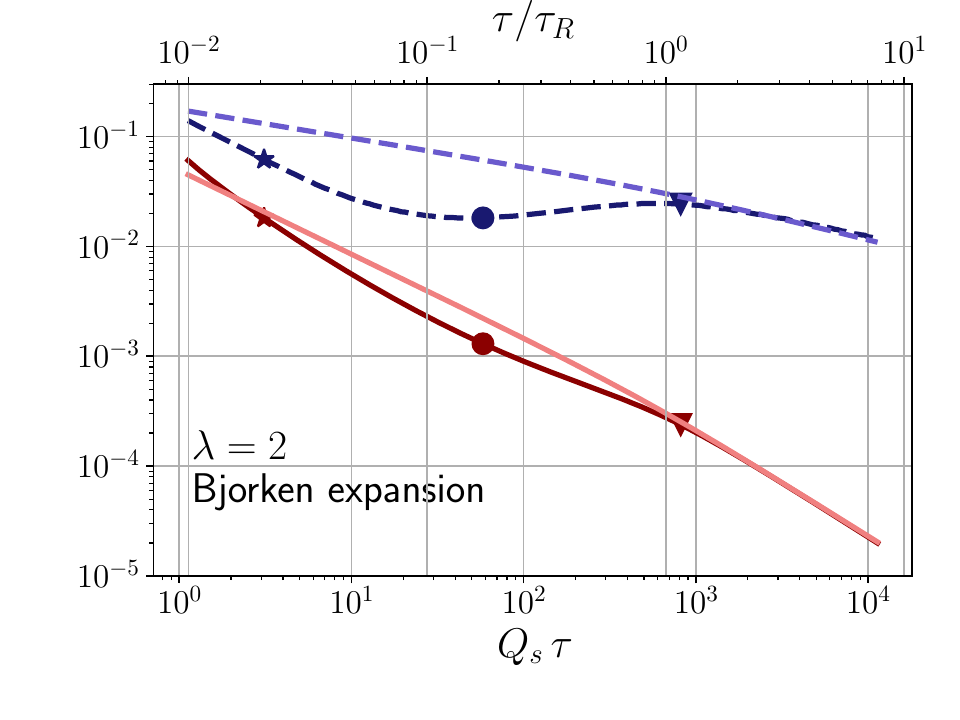}
    }
    \centerline{
        \includegraphics[width=0.5\linewidth, trim={6.5cm 6.5cm 0cm 0cm}, clip]{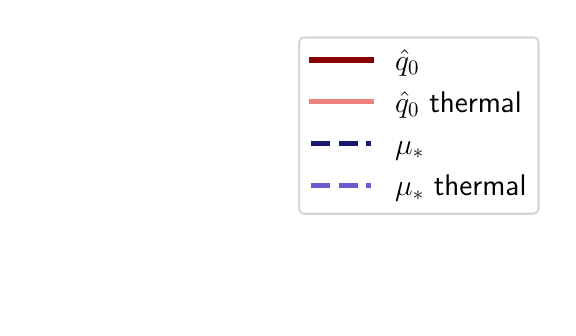}
        \includegraphics[width=0.5\linewidth]{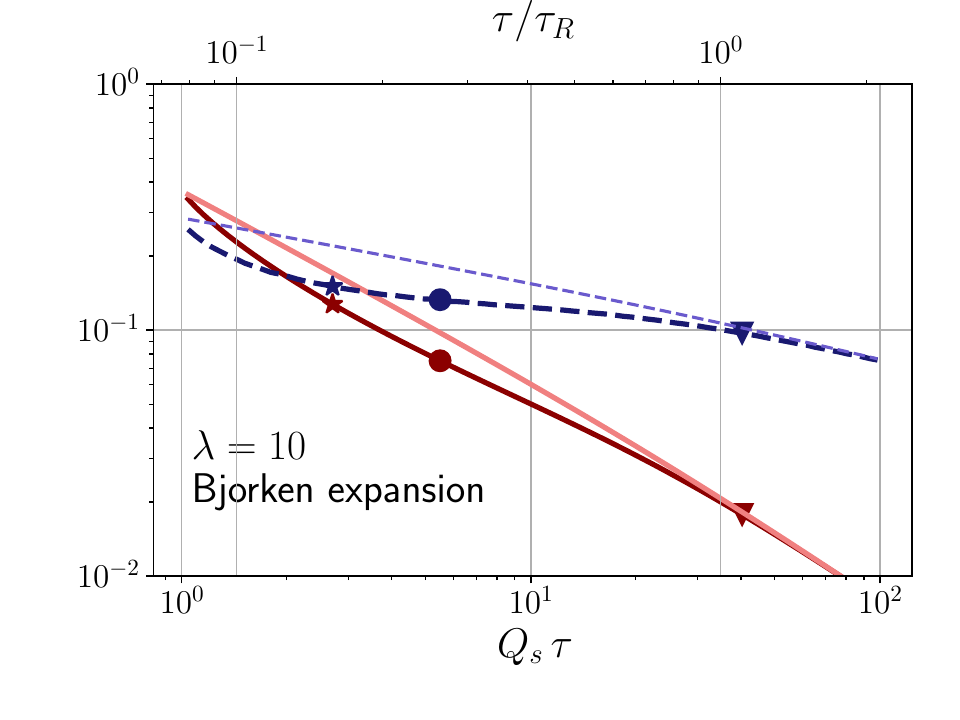}
    }
    \caption{
    Time evolution of the bare jet quenching parameter $\hat q_0$, and screening mass $\mu_\ast$ for the systems considered in this paper, as extracted from the EKT simulations. In the left column, we show the isotropic systems, where the quantities are depicted in units of $T$ (underoccupied system, top panel) and $Q$ (overoccupied system, bottom panel). In the right column, we show $\hat q_0/Q_s^3$ and $\mu_\ast/Q_s$ for the expanding system in units of the saturation momentum $Q_s$ for various couplings.
    }
    \label{fig:qhat_plots}
\end{figure*}

In Fig.~\ref{fig:qhat_plots}, we show the evolution of $\hat q_0$ and $\mu_\ast$ for the nonequilibrium system, defined through $a$ and $b$ in Eq.~\eqref{eq:qhat_parametrization} and obtained from QCD kinetic theory simulations. These nonequilibrium results are compared with the equilibrium case computed for a thermal system with the same energy density (Landau matching condition). For isotropic, nonexpanding systems (left column), the thermal reference yields constant $\hat q_0$ and $\mu_\ast$, whereas for expanding systems the continuously decreasing energy density also reduces the thermal values of $\hat q_0$ and $\mu_\ast$ (right column). In practice, the effective temperature of the thermal system is obtained from the nonequilibrium energy density of gluons 
\begin{align}
     \Teps(t) = \left(\frac{30\varepsilon(t)}{\nu \pi^2}\right)^{1/4}. \label{eq:Teps}
\end{align}
This temperature is then used for the corresponding equilibrium values \cite{Arnold:2008vd}
\begin{subequations}\label{eq:ab_thermal}
\begin{align}
    a(t) &= \Teps^3(t) \frac{N_CC_Rg^4\zeta(3)}{\pi^3}\, ,\\
    b(t) &=C_R\frac{g^4\Teps^3(t)}{\pi^3}N_C \Bigg(-\zeta(3)\ln\frac{Q_s}{m_D} -\frac{\sigma_+}{2\pi}\nn& + (\zeta(2)-\zeta(3))\left[\ln\frac{\Teps}{m_D}+\frac{1}{2}-\gamma_E+\ln 2\right]\Bigg) \, ,
\end{align}
\end{subequations}
with $m_D^2(t)=g^2\Teps^2(t)\frac{N_C}{3}$, 
and  where 
$\gamma_E\approx 0.57722(\dots)$ is the Euler-Mascheroni constant, and $\sigma_+=\sum_{k=1}^\infty \frac{\log[(k-1)!]}{k^3}=0.386043817(\dots)$. 
Thus, we finally find
 \begin{align}
    \mu_\ast^2&=\frac{m_D^2}{4}\left(\frac{m_D}{2\Teps(t)}\right)^{2\zeta(2)/\zeta(3)-2}\\
    &\times\exp\left[2\gamma_E-2-\left(\frac{\zeta(2)}{\zeta(3)}-1\right)(1-2\gamma_E)+\frac{\sigma_+}{\pi\zeta(3)}\right]\nonumber\,.
\end{align}
Note that in~\cite{Arnold:2008vd} one assumes a large scale separation $m_D\ll \Teps$, which is true at weak couplings, for the derivation of the thermal values in Eq.~\eqref{eq:ab_thermal} starting from Eq.~\eqref{eq:qhat_general}.
It was noticed in~\cite{Boguslavski:2023waw} that for couplings that are not sufficiently small,
the approximations of Ref.~\cite{Arnold:2008vd} lead to a different result than when numerically evaluating Eq.~\eqref{eq:qhat_general} with a thermal distribution function.
To ensure that our numerically extracted
$\hat q$ approaches the appropriate thermal 
limit of Eq.~\eqref{eq:qhat_general} at late times, we take the equilibrium value%
\footnote{In particular, we use the value of $\tilde b$ from Table III of \cite{Boguslavski:2023waw}, which differs from the parameter $b$ used here, as in~\cite{Boguslavski:2023waw}, the large $\Lambda_\perp$ behavior is parametrized using $\log \Lambda_\perp/m_D$ instead of $\log \Lambda_\perp/Q_s$, {\it c.f.} Eq.~\eqref{eq:qhat_parametrization}.} 
of $b$ from Table III of \cite{Boguslavski:2023waw} for $\lambda \geq 2$.

Additionally, we will 
compare our nonequilibrium results to an equivalent system with constant $\qhat_0$ and $\mu_\ast$, which can be thought of as a static QGP brick. 
Note that there 
is no formal mapping between an expanding and a static system. Thus, this comparison should be 
understood at a 
phenomenological level, as is commonly done in the literature, see e.g.~\cite{Blok:2024tyb,Arleo:2022shs,Caucal:2020uic} for recent examples. For $\hat q_0^{\mathrm{static}}$, we adopt a mapping procedure analogous to that in~\cite{Salgado:2002cd}, which connects an expanding medium to a static one by enforcing 
\begin{align}
    \hat q_0^{\mathrm{static}}=\frac{2}{L^2}\int_{t_0}^{t_0+L}\dd t\, (t-t_0)\,\hat q_0(t)\, ,\label{eq:qhat0-static}
\end{align}
such that the characteristic energy scales of the emitted gluons with formation time comparable to the system size is preserved, 
\begin{align}
\omega_c = \frac{1}{2}\qhat_0^{\mathrm{static}}L^2. \label{eq:omegac}
\end{align}
We will frequently denote the energy of the emitted gluon relative to this characteristic gluon frequency $\omega_c$.
We further choose $\mu_\ast^{\mathrm{static}}$ such that the transverse-momentum-broadening matching scale, see Eq.~\eqref{eq:Qb_final}, 
\begin{align*}
    Q_b^2=\int_{t_0}^{t_0+L}\dd t\, \hat q_0(t)\log\frac{Q_b^2}{\mu_\ast^2(t)}\, ,
\end{align*}
is the same for both the nonequilibrium and static systems. In practice, we first determine $Q_b$ from Eq.~\eqref{eq:Qb_final} for the nonequilibrium system, and then set
\begin{align}
    \mu_\ast^{\mathrm{static}} = Q_b e^{-\frac{Q_b^2}{2L \hat q_0^{\mathrm{static}}}} \, .\label{eq:mustar-static}
\end{align}

Let us now discuss the numerical results of the kinetic theory simulations $\hat q_0(t)$ and $\mu_\ast(t)$, which serve as input for
the spectrum, see Fig.~\ref{fig:qhat_plots}. 
The left column shows the isotropic, non-expanding systems. The upper panel corresponds to the initially underoccupied case, where both the bare quenching parameter $\hat q_0$ and the screening mass $\mu_\ast$ approach their thermal values from below. Physically, this configuration represents a number of hard particles with momentum $Q$, which generate a soft thermal bath through soft gluon radiation. The system thermalizes once the hard gluons have transferred all their energy to this bath. In contrast, the initially overoccupied system shown in the central left panel exhibits the opposite behavior: $\hat q_0$ approaches its thermal value from above, reflecting the larger initial number density of gluons. In this case, the screening mass $\mu_\ast$ remains approximately constant at its thermal value. In these isotropic simulations, the energy density is constant, implying a fixed temperature and, thus, leading to constant values of $\hat q_0$ and $\mu_\ast$.

The results for the expanding system are shown in the right column of Fig.~\ref{fig:qhat_plots}. These correspond to systems undergoing bottom-up thermalization \cite{Baier:2000sb}, which occurs in several distinct stages. Following previous works, see e.g. \cite{Boguslavski:2023alu}, we include markers that visually indicate the approximate positions of these individual stages. Initially, the system is overoccupied (see the initial condition $\sim 1/\lambda$ in Eq.~\eqref{eq:initial_cond_expanding}), and the first stage therefore resembles the initially overoccupied isotropic case. As the evolution continues, the occupancy rapidly decreases while the expansion drives the system toward increasing anisotropy. We place the star marker at the point where the occupancy $\langle p f\rangle/\langle p\rangle$ drops below $1/\lambda$, indicating the transition out of the classical overoccupied regime. This behavior is particularly evident in the panels for weaker couplings, where the initial $\hat q_0$ exceeds its equilibrium value.

After the star marker, the expansion drives the system into the underoccupied regime, while it remains highly anisotropic. In this stage, the occupancy continues to decrease until it reaches a minimum, indicated by the circle marker. Beyond this point, thermalization proceeds similarly to the initially underoccupied isotropic case shown in the upper left panel. Physically, the remaining hard particles become quenched as they traverse the soft thermal bath formed in the earlier stages. As in the isotropic underoccupied case, the nonequilibrium values of the bare quenching parameter $\hat q_0$ and the screening mass $\mu_\ast$ lie below their thermal counterparts and approach them from below. The triangle marker denotes the point where the system has nearly isotropized, defined here by the condition $P_L/P_T=0.5$. After this point, $\hat q_0$ and $\mu_\ast$ are in excellent agreement with their thermal values.
A somewhat special case is the phenomenologically more relevant coupling $\lambda=10$. Here, the overoccupied stage appears to be absent, as the initial condition already lies close to the underoccupied regime.

Comparing simulations with different couplings, we expect $\hat q_0$ to scale as $\hat q_0\sim \lambda^{5/4}$ initially. This arises from a combination of two effects: when neglecting corrections coming from the screening mass, the matrix element scales as $|\mathcal M|^2\sim \lambda^2$. Additionally, the initial energy density is proportional to $1/\lambda$ due to our initial condition \eqref{eq:initial_cond_expanding}, which leads to $T\sim \lambda^{-1/4}$. 
Together, they yield indeed $\hat q_0 \sim \lambda^2 T^3 \sim \lambda^{5/4}$.

In general, for the expanding systems, the thermal values
$\qhat_0^{\mathrm{thermal}}$ and $\mu_\ast^{\mathrm{thermal}}$ are larger than their nonequilibrium counterparts after the overoccupied regime (i.e. after the star marker), and thus throughout most of the evolution. 
This has two competing effects on the effective jet quenching parameter $\qhat_r^{\mathrm{thermal}}$. 
The larger $\qhat_0^{\mathrm{thermal}}$ enhances $\qhat_r^{\mathrm{thermal}}$, leading to stronger quenching in the thermal system. 
In contrast, the larger $\mu_\ast^{\mathrm{thermal}}$ enters the denominator of the logarithm in $\qhat_r^{\mathrm{thermal}}$, 
slightly reducing its effective value and partially compensating the enhancement from $\qhat_0^{\mathrm{thermal}}$.

While we normalize all dimensionful quantities in units of the saturation momentum $Q_s$ for the expanding case, or the final temperature $T$ or typical momentum $Q$ in the underoccupied and overoccupied cases, we also depict additional time scales. For the overoccupied case, the top scale shows time in units of the equilibrium temperature. For the expanding systems, we additionally show time in units of the relaxation time $\tau_R$,
\begin{align}
    \tau_R=\frac{4\pi\eta/s}{\Teps(\tau)},
\end{align}
with the time-dependent temperature defined in Eq.~\eqref{eq:Teps} and the specific shear viscosity $\eta/s$ extracted in~\cite{Boguslavski:2024kbd}.
This time scale governs the relaxation to equilibrium in first-order hydrodynamics.

\subsection{On the applicability of the IOE in out-of-equilibrium matter}
Having discussed the two main theoretical ingredients of our calculation---the soft induced gluon spectrum within the IOE framework and the extraction of the jet transport coefficient and screening mass from EKT simulations---we now briefly comment on the validity of applying the IOE to out-of-equilibrium matter.

QCD kinetic theory, as formulated in the seminal work by Arnold, Moore, and Yaffe (AMY)~\cite{Arnold:2002zm}, includes both elastic and inelastic processes. The former govern particle diffusion, while the latter describe gluon production sourced by \textit{hard} modes in the plasma~\cite{Arnold:2002ja}. Importantly, the AMY framework incorporates both incoherent gluon emission, relevant in the low-energy Bethe-Heitler regime, and quantum-coherence phenomena associated with the QCD LPM effect. Consequently, the AMY approach provides a unified description of thermal QCD plasmas and the radiation pattern of jets in HICs. 

Although broadly applicable, the AMY formalism relies on several assumptions about the properties of the medium under consideration, which are neatly summarized in~\cite{Arnold:2002zm}. In particular, plasma instabilities are not included, which are generically present in anisotropic systems \cite{Mrowczynski:1993qm, Romatschke:2003ms, Mrowczynski:2016etf}, and thus isotropic screening prescriptions are used. Therefore, while AMY can be extended to out-of-equilibrium settings---where an effective temperature emerges and much of the structure of the key results is preserved---its direct use in anisotropic systems requires careful and detailed reconsideration.

Revisiting the assumptions underlying the BDMPS-Z description, which forms the basis of the IOE, one finds that they essentially coincide with those adopted in AMY. In particular, the medium is assumed to be isotropic in the transverse directions. This observation leads to the following two conclusions: \textbf{1)} extending the BDMPS-Z/IOE framework to fully anisotropic systems only makes sense if a corresponding generalization is also implemented on the EKT side, \textbf{2)} otherwise, the two frameworks would become conceptually inconsistent. Neglecting such effects on both sides, however, renders the entire construction self-consistent. In this sense, the IOE can be consistently applied to an out-of-equilibrium medium described by the EKT, provided that anisotropic effects and plasma instabilities are not taken into account. We note that recent works have introduced an extension of the BDMPS-Z formalism to anisotropic backgrounds, see e.g. \cite{Barata:2023qds}, but a detailed study of these effects in out-of-equilibrium media is left for future work.

\section{Numerical results for the gluon spectrum}\label{sec:numerics}
In this section, we discuss the numerical results for the evaluation of the radiation spectrum in the three scenarios introduced above for the bulk matter. We further compare our results with those obtained using the harmonic approximation for the scattering potential.

\subsection{Results for isotropic nonexpanding plasmas}
\begin{figure}
    \centering
    \includegraphics[width=\linewidth]{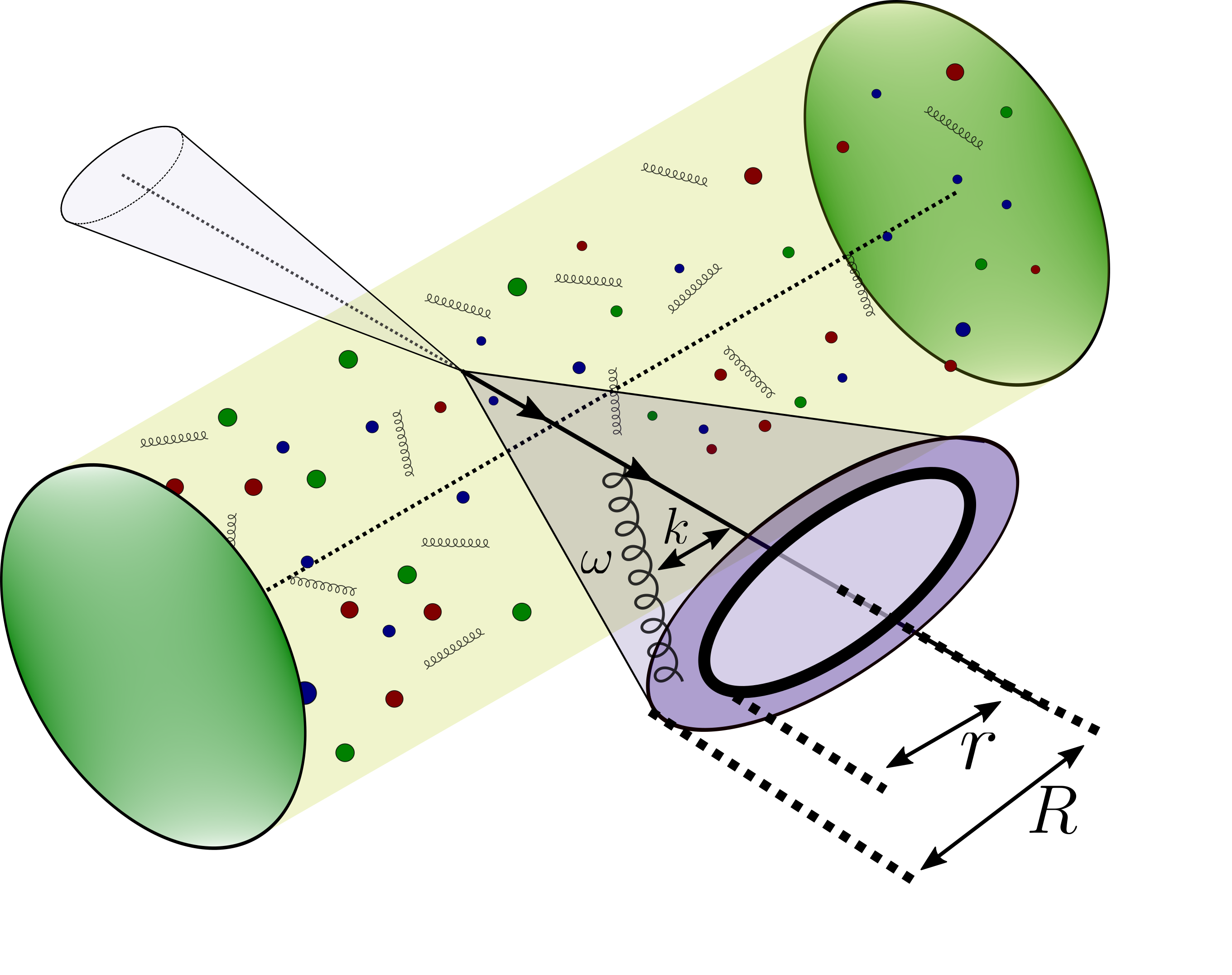}
    \caption{Illustration of the jet shape at leading order in $\alpha_s$: the probability of energy deposited in the \emph{annulus} of radius $r$ and $R$.}
    \label{fig:jetshape}
\end{figure}

\begin{figure*}
    \centering
    \centerline{
        \includegraphics[width=0.33\linewidth]{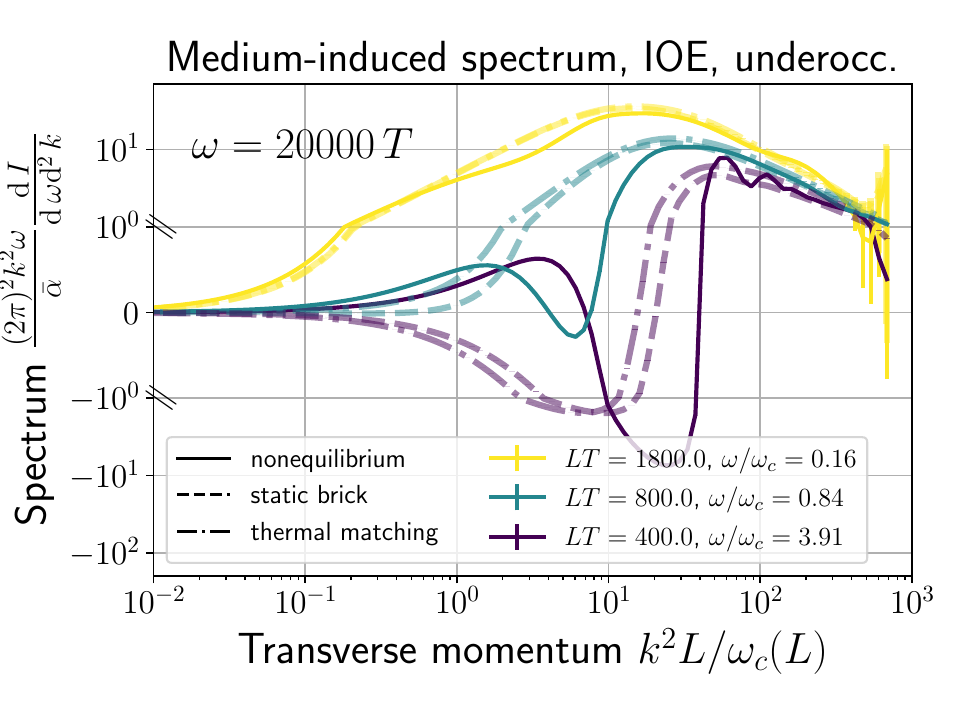}
        \includegraphics[width=0.33\linewidth]{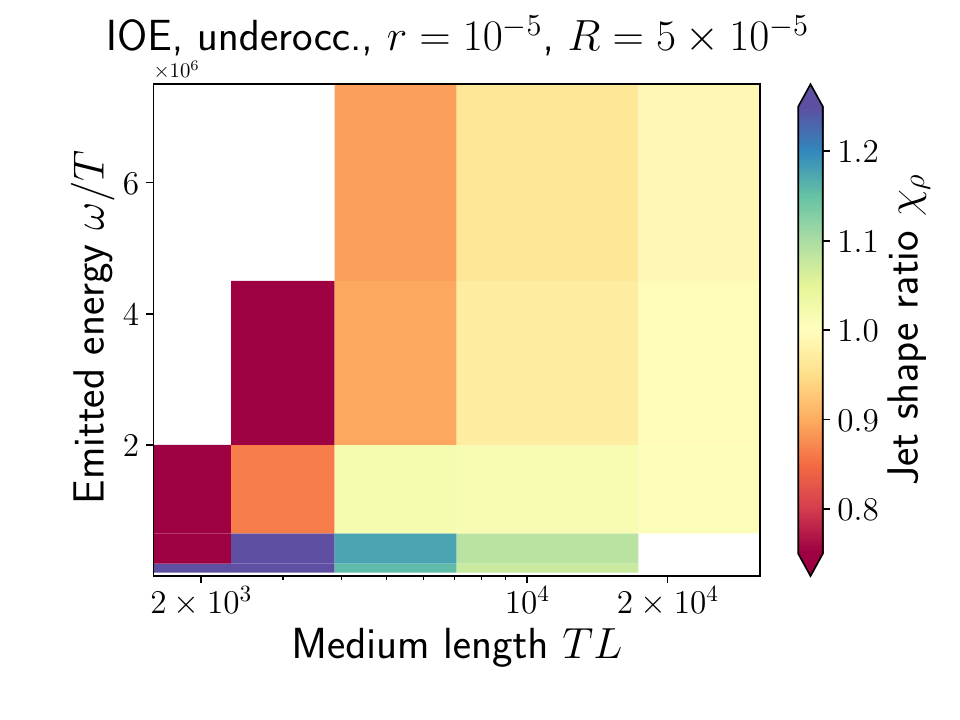}
        \includegraphics[width=0.33\linewidth]{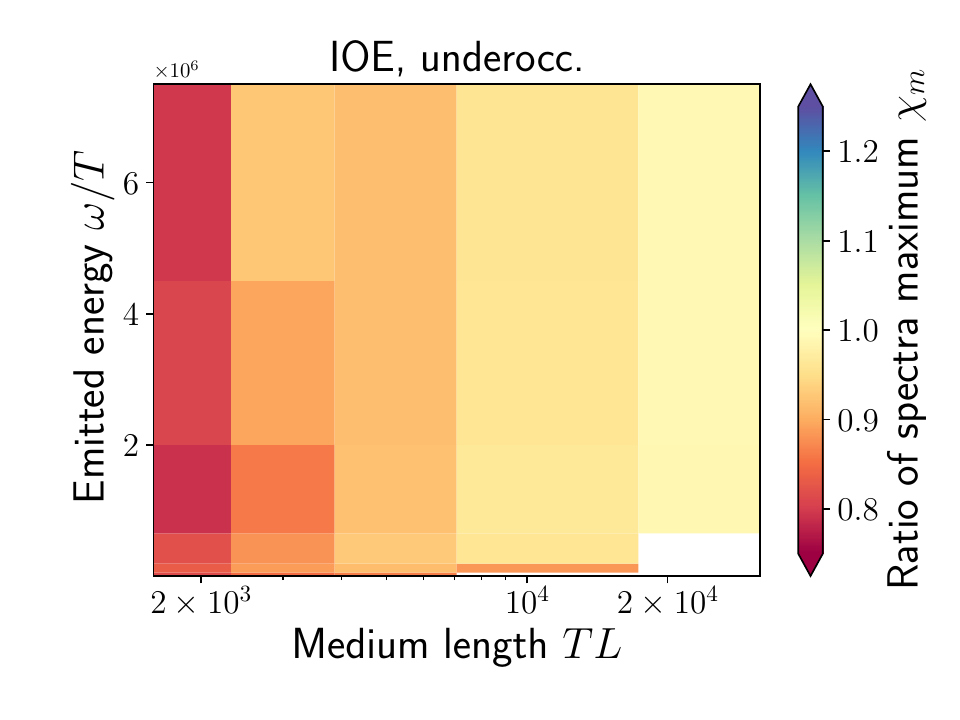}
    }
    \centerline{
        \includegraphics[width=0.33\linewidth]{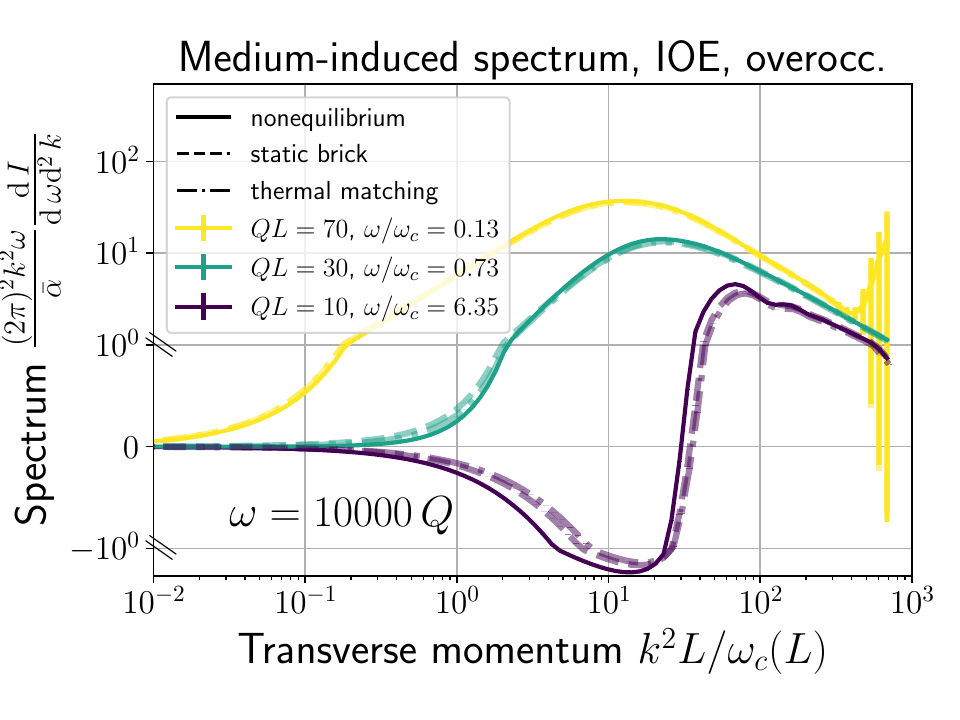}
        \includegraphics[width=0.33\linewidth]{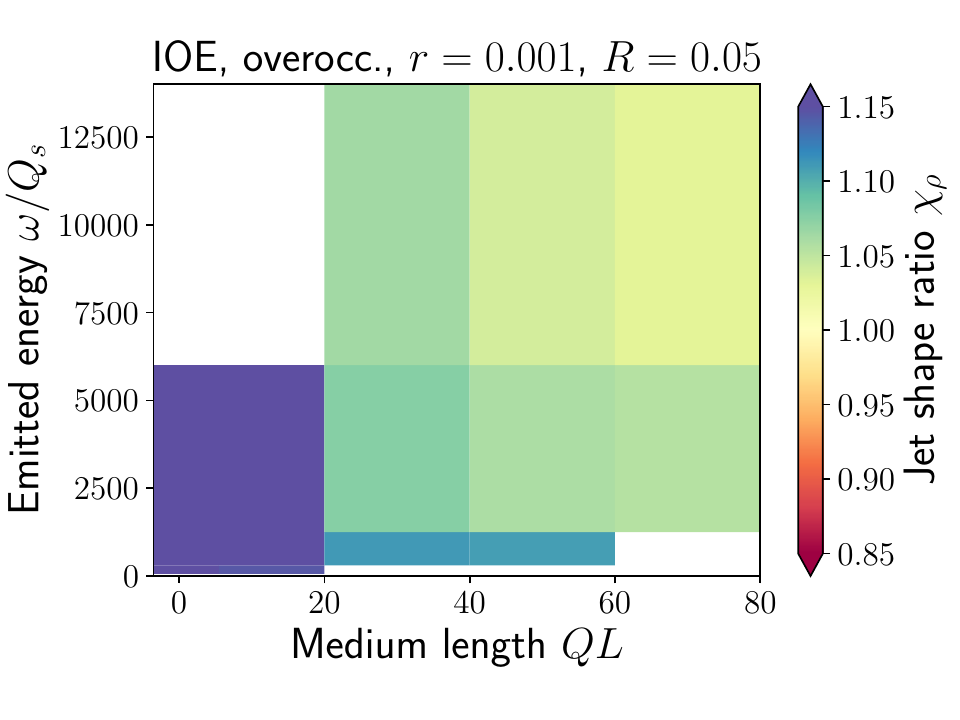}
        \includegraphics[width=0.33\linewidth]{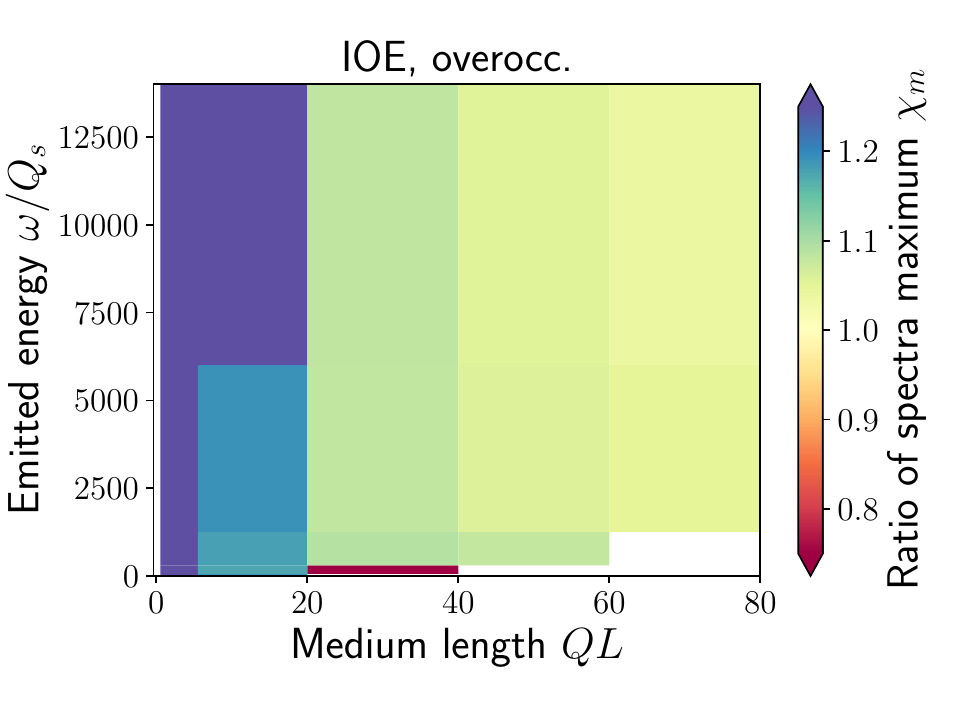}
    }
    \caption{
    Medium-induced gluon spectrum (left column) for an isotropic initially underoccupied (top row) and overoccupied (bottom row) system. The center and right column show the ratio observables $\chi_i$ defined in Eqs.~\eqref{eq:ratio-jetshape} and \eqref{eq:ratio-maxima}.
    }
    \label{fig:results-iso}
\end{figure*}

\begin{figure*}
    \centering
    \centerline{
        \includegraphics[width=0.33\linewidth]{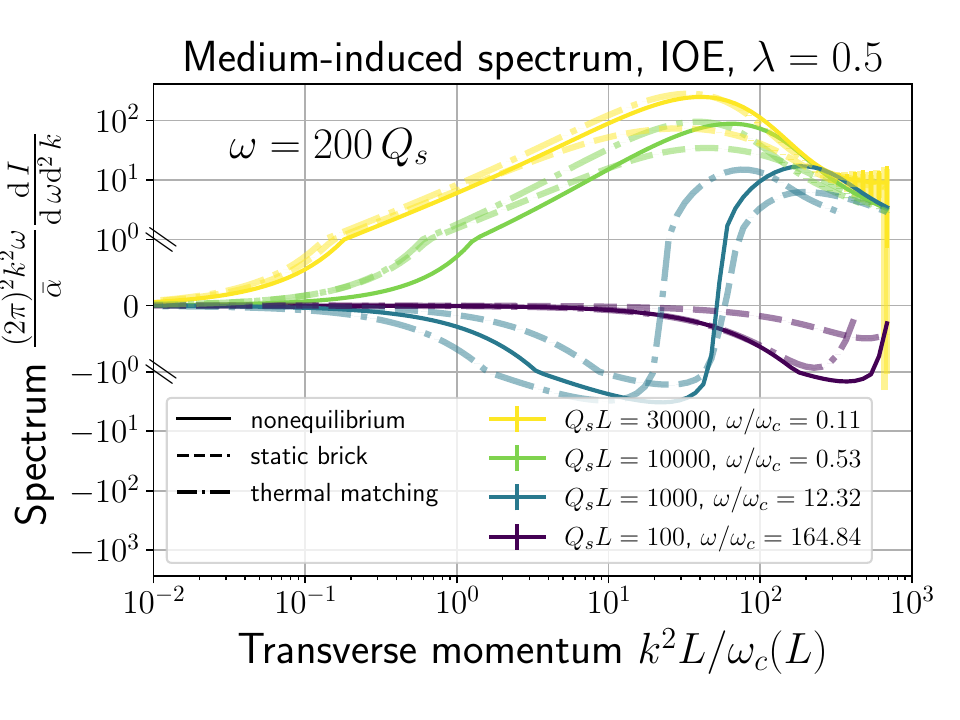}
        \includegraphics[width=0.33\linewidth]{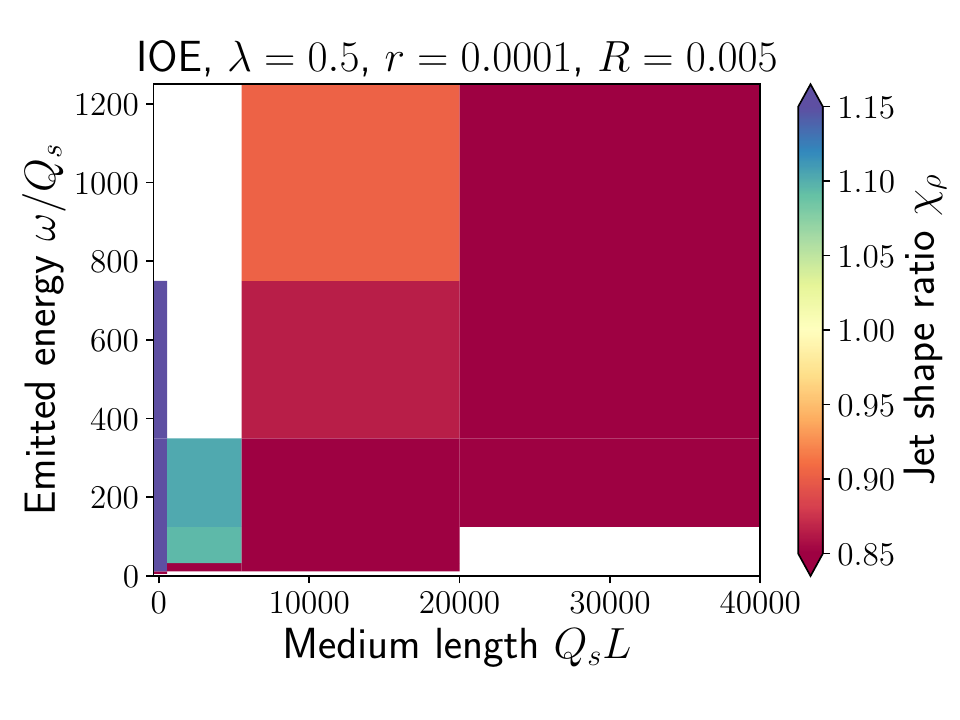}
        \includegraphics[width=0.33\linewidth]{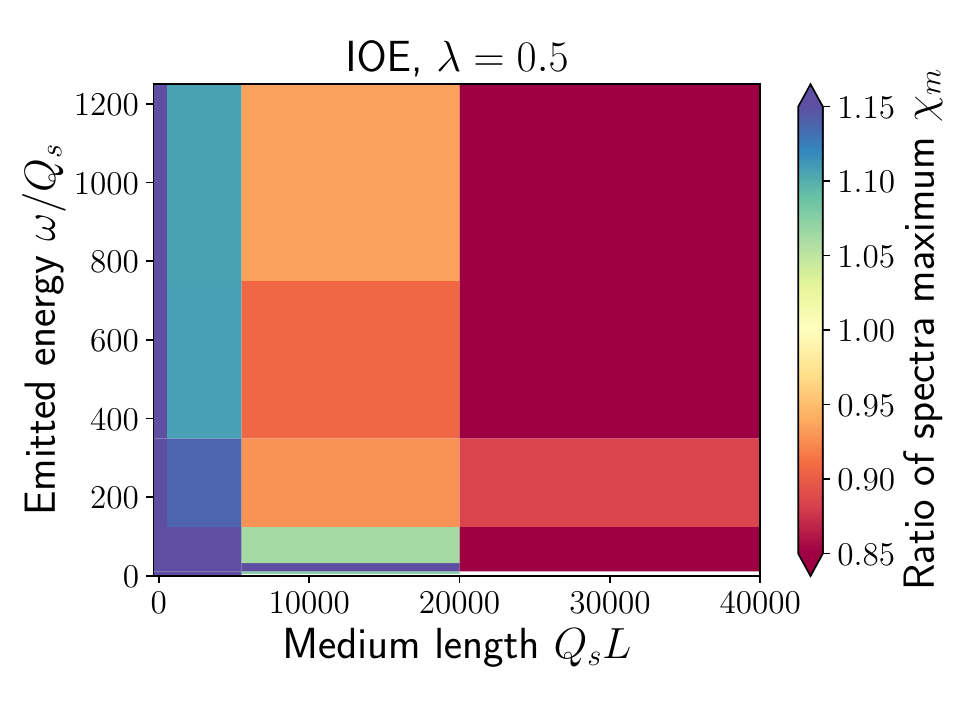}
    }
    \centerline{
        \includegraphics[width=0.33\linewidth]{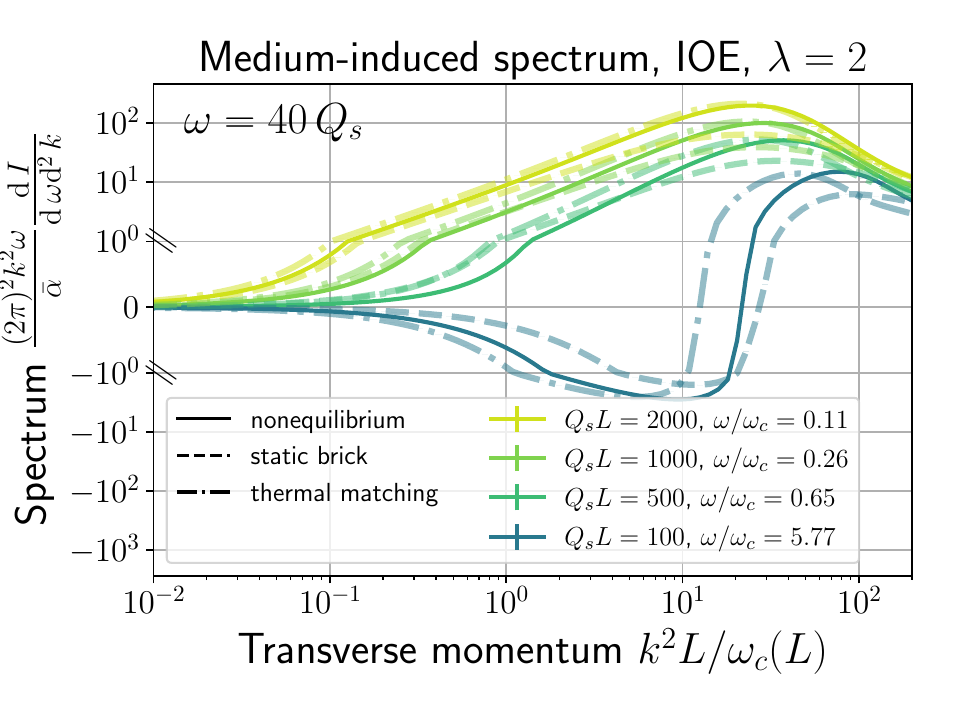}
        \includegraphics[width=0.33\linewidth]{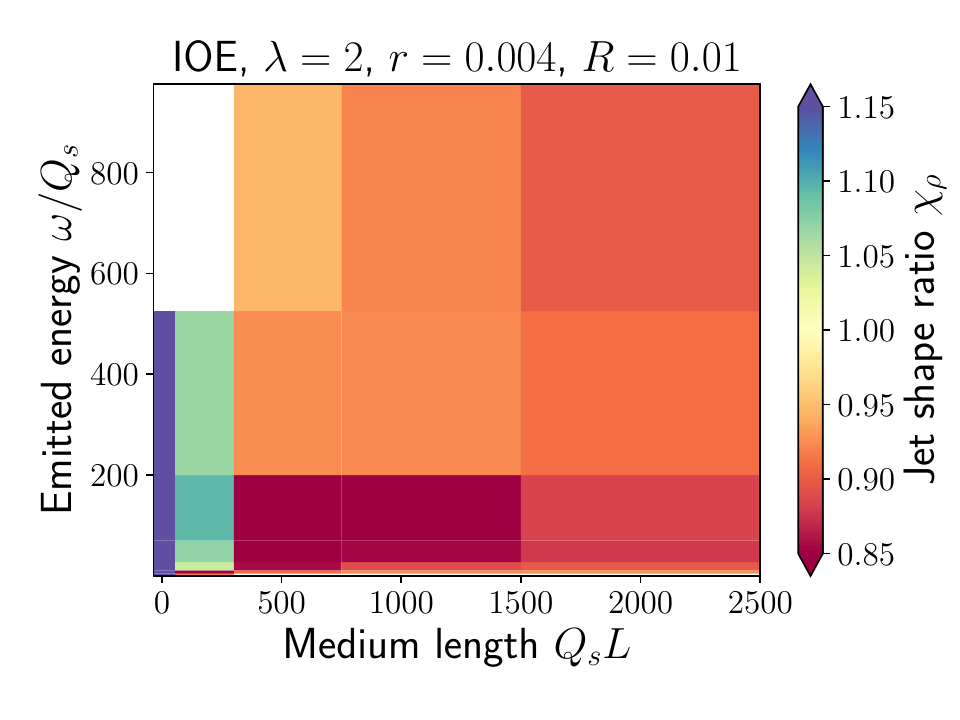}
        \includegraphics[width=0.33\linewidth]{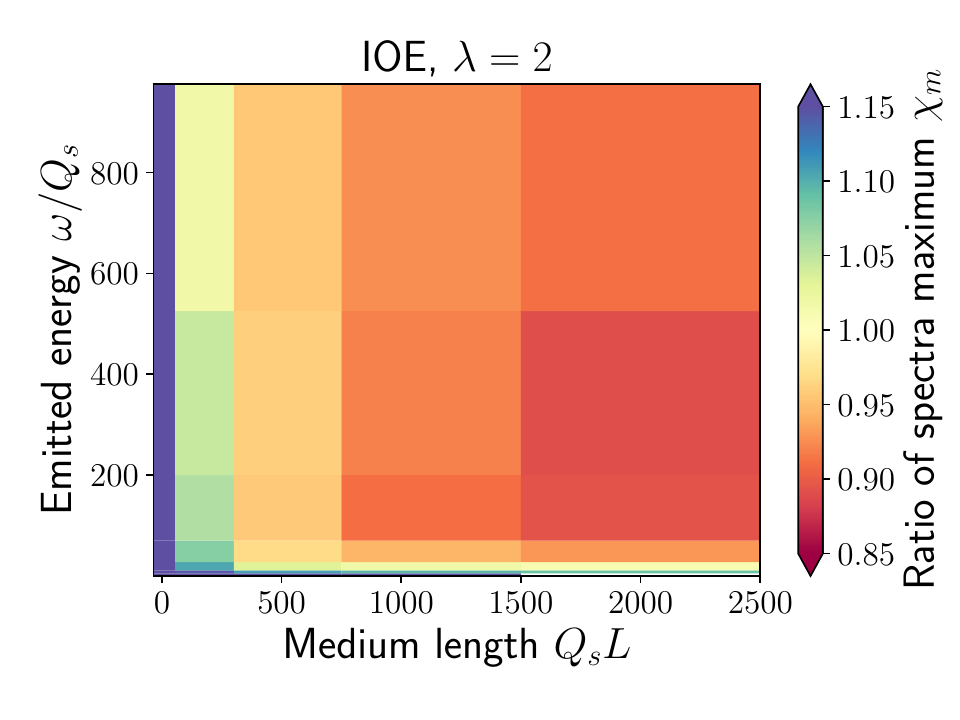}
    }
    \centerline{
        \includegraphics[width=0.33\linewidth]{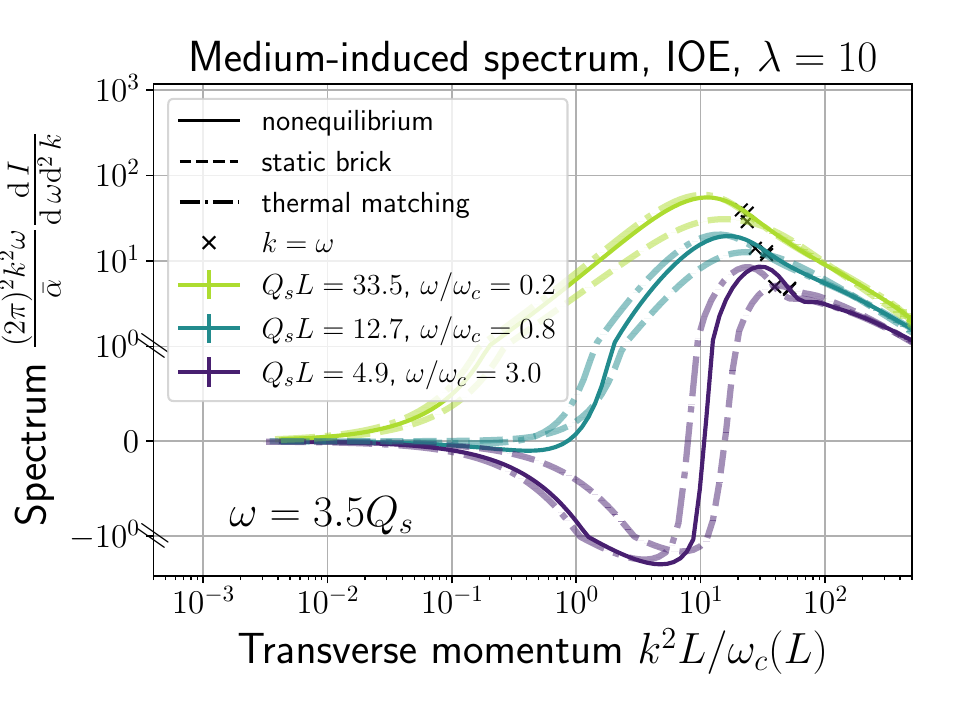}
        \includegraphics[width=0.33\linewidth]{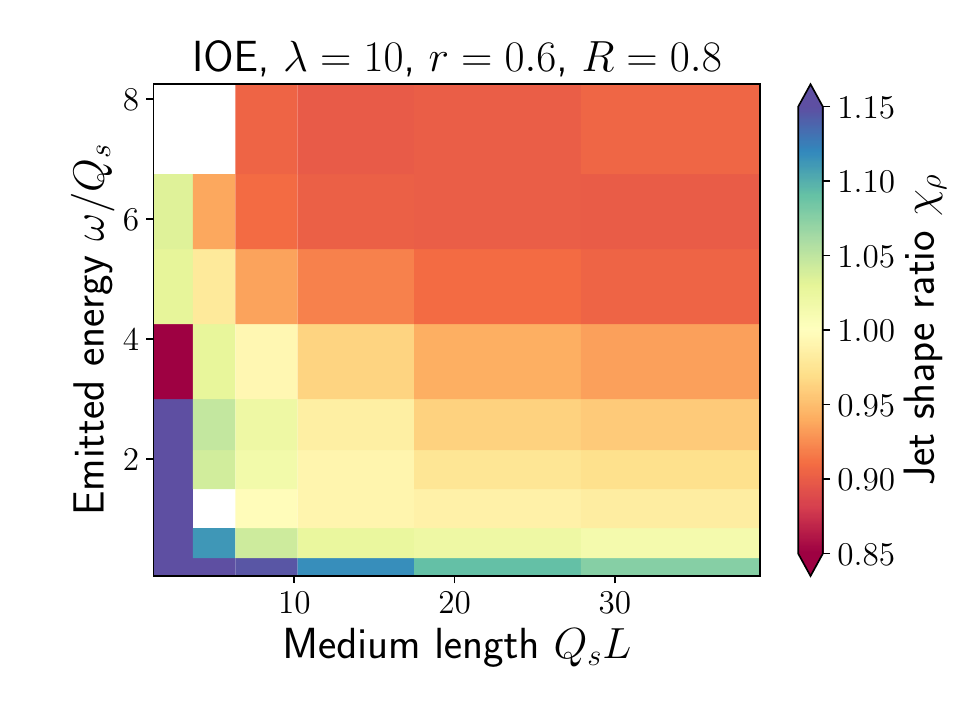}
        \includegraphics[width=0.33\linewidth]{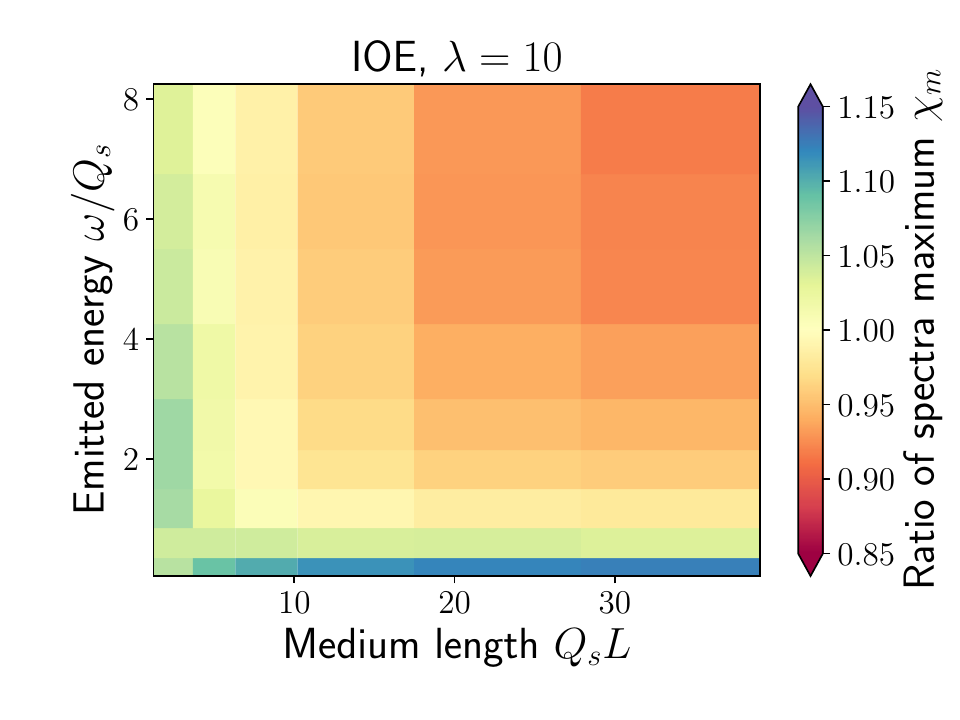}
    }
    \caption{
    Medium-induced gluon spectrum (left column) for Bjorken expanding matter. In the middle and right columns, we show the ratio observables $\chi_i$ defined in Eqs.~\eqref{eq:ratio-jetshape} and \eqref{eq:ratio-maxima}.
    }
    \label{fig:bjorken}
\end{figure*}

We start by discussing the numerical results for an isotropic, non-expanding plasma. We compare plasmas of different lengths, which are obtained by setting $\qhat(t)\equiv 0$ for $t>L+\tmin$, while taking $\qhat(t)$ from the EKT simulation shown in the left panels of Fig.~\ref{fig:qhat_plots}. The evolution of $\qhat$ is then effectively
\begin{align}
	\qhat(t)&=\begin{cases}
		\qhat^{\mathrm{sim}}(t), &  \tmin \leq t \leq \tmax\\
		0, & t > \tmax
	\end{cases}\label{eq:effective-qhat-evolution},
\end{align}
with $\tmax=L+\tmin$.

The resulting spectrum $\frac{\dd{I}}{\dd\omega\dd^2k}$ is obtained by numerically evaluating the integrals of Eqs.~\eqref{eq:lo-spectrum} and \eqref{eq:nlo-spectrum}, with the functions $C$ and $S$ being solutions to the differential equation \eqref{eq:differential-equation}, which are also obtained numerically. The spectrum itself represents the probability of emitting a gluon with energy $\omega$ and transverse momentum $k$.
The results for the spectrum at a fixed gluon energy $\omega$, normalized to the corresponding characteristic frequency $\omega_c$ from Eq.~\eqref{eq:omegac}, are shown in the left column of Fig.~\ref{fig:results-iso}, where different medium lengths are indicated by different colors. The upper row displays the results for an initially under-occupied system, while the lower row shows those of an initially over-occupied plasma. The $y$ axis in the left column uses a symmetric logarithmic scale, with a linear region around $0$ and a logarithmic region symmetrically starting at the point indicated by two lines, signaling the scale change. Importantly, while the nonequilibrium and static curves have the same $\omega_c$, the equilibrium curve is constructed from the energy density, and thus its characteristic frequency $\omega_c$ slightly differs. All curves in the plot have the same gluon energy $\omega$. The ratio $\omega/\omega_c$ given in the legend refers to the ratio in the nonequilibrium system.
For each length, we show three distinct curves: the solid line depicts the spectrum obtained from the nonequilibrium evolution of $\qhat_0$ and $\mu_\ast$, as shown in Fig.~\ref{fig:qhat_plots}, the dash-dotted curves represent the evolution for the thermally matched background, and the dashed curves show the results for a static brick, obtained via the matching procedure explained in the previous section, {\it i.e.} see Eqs.~\eqref{eq:qhat0-static} and \eqref{eq:mustar-static}.

We observe that in general, for a fixed gluon energy $\omega$, changing the length leads to the most important qualitative and quantitative changes of the spectrum. In particular, for the simpler harmonic-oscillator approximation that we will compare to further below, the shape of the spectrum is mainly controlled by the ratio $\omega/\omega_c$ \cite{Salgado:2003gb}.
Moreover, for the nonexpanding systems, the spectra obtained using thermal matching closely follow those computed with the full nonequilibrium evolution of $\qhat$, particularly in the initially over-occupied case. This indicates that the thermal reference---an equilibrium medium with the same instantaneous energy density---provides a reasonable approximation to the nonequilibrium result in nonexpanding systems, although the differences remain noticeable. In the isotropic case, the static-brick matching
provides similar results as the thermal matching.
For the initially over-occupied system, in the vicinity of the dominant peak, the thermal values lie mostly below the nonequilibrium ones, while the opposite trend is observed for the initially under-occupied case.%
\footnote{
Note that the fact that the spectrum becomes negative for some values of the parameters is not unphysical; it represents a suppression of mediun-induced radiation with respect to vacuum emissions, and appears to occur for $\omega>\omega_c$, where $\omega_c$ is the frequency for which the formation time is parametrically of the same order as the medium size. In general, the formation time is also dependent on the transverse momentum, and the relation becomes more complicated.
This feature is known to emerge even in equilibrium matter, e.g., appearing in the harmonic approximation, as is discussed in Section \ref{sec:ioe-ho-comparison}.
}

To better quantify the differences between the spectra, we define two auxiliary ratios, shown in the central and right columns of Fig.~\ref{fig:results-iso}. We first refer to the jet shape, see e.g.~\cite{Chien:2015hda, Milhano:2022kzx}, which characterizes the energy deposited by the emitted gluon between two angular scales or jet radii  $r$ and  $R$ (see Fig.~\ref{fig:jetshape}). It is defined as
\begin{align}
    \dv[\rho(r,R)]{\omega}=\int_{r\omega}^{R\omega}\dd^2{\vb \k} \,\omega \frac{\dd I}{\dd\omega\dd^2{\vb \k}} \, .
\end{align}
We then introduce the jet-shape ratio, comparing the nonequilibrium and equilibrium results,
\begin{align}
    \chi_\rho(\omega,L; r,R)=\frac{\dv[\rho^\mathrm{noneq}(r,R)]{\omega}}{\dv[\rho^{\mathrm{eq}}(r,R)]{\omega}} \, , \label{eq:ratio-jetshape}
\end{align}
and show it in the central column of Fig.~\ref{fig:jetshape}. 
The right column, in turn, shows the ratio of the maximal values of the spectra,
\begin{align}
    \chi_m(\omega, L)=\frac{\max_k\frac{k^2\dd{I^\mathrm{noneq}}}{\dd\omega \dd^2{\vb k}}}{\max_k\frac{k^2\dd{I^\mathrm{eq}}}{\dd\omega \dd^2{\vb k}}}. \label{eq:ratio-maxima}
\end{align}
Both ratios quantify the deviation of the nonequilibrium spectrum from its equilibrium counterpart, with $\chi_i=1$ corresponding to exact agreement.

Confirming our previous observations about the behavior in the vicinity of the peak of the spectrum, for the initially under-occupied case (upper row), these ratios are below $1$ for smaller media and lower gluon energies $\omega$, while approaching $\chi_i=1$ as $L$ increases. This behavior is expected, as the initially nonequilibrium plasma gradually thermalizes, and a jet produced in the early nonequilibrium stage will traverse a medium that remains closer to equilibrium for longer as $L$ increases. Similarly, for the initially over-occupied setup, the largest deviations from equilibrium occur for smaller systems, which are further away from equilibrium. In this case, the ratio is larger than $1$ and it approaches the equilibrium value from above for larger $L$.

\subsection{Systems with Bjorken expansion}

\begin{figure}
    \centerline{
        \includegraphics[width=.85\columnwidth]{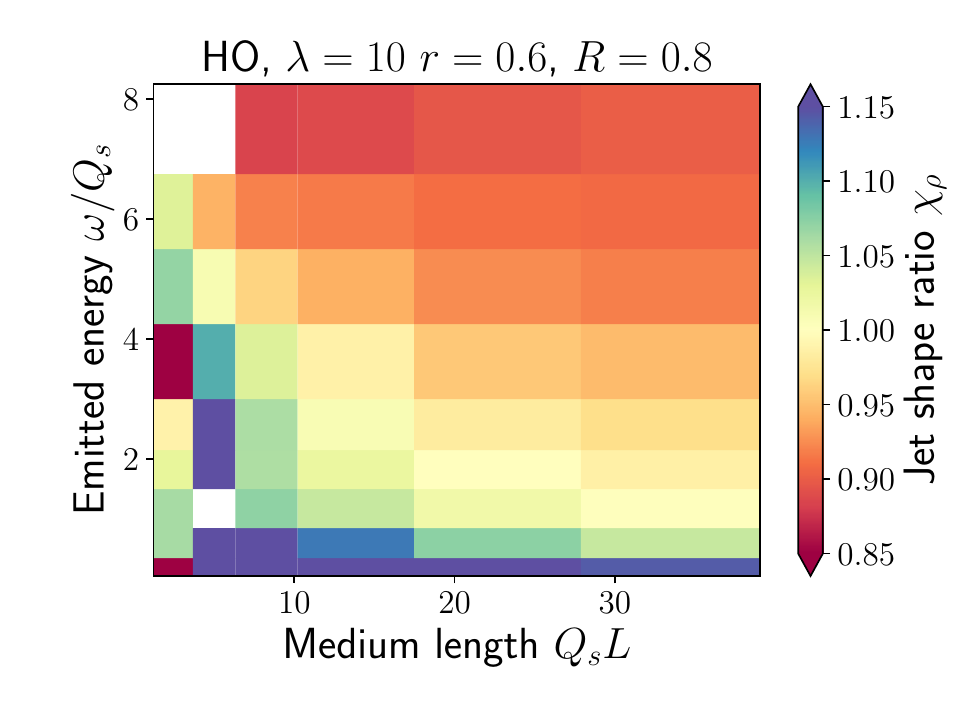}
    }
    \centerline{
        \includegraphics[width=.85\columnwidth]{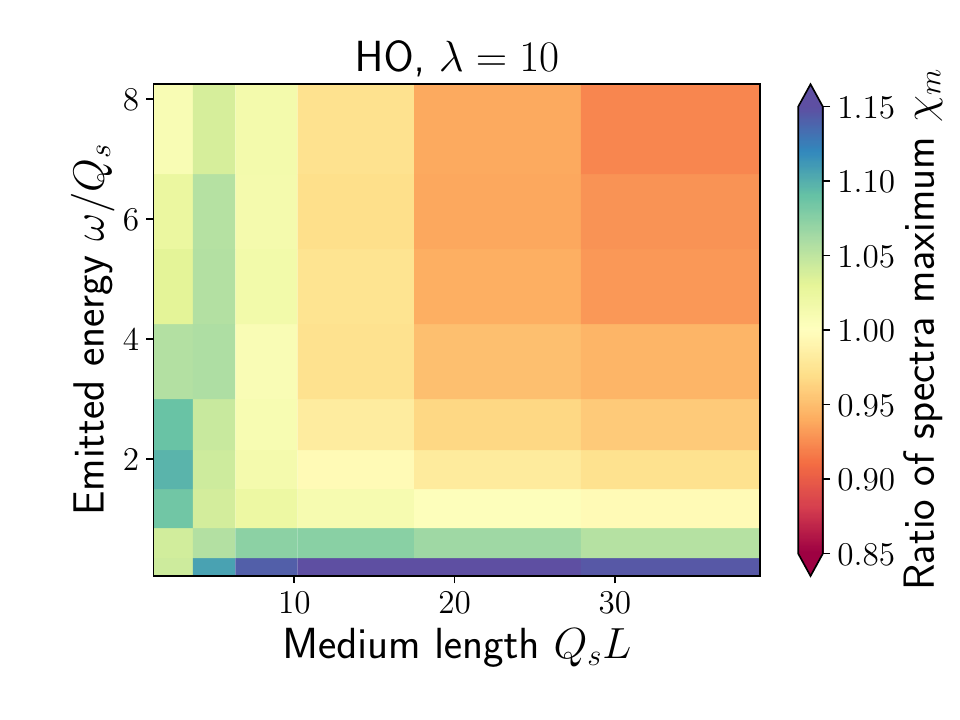}
    }
    \caption{The ratio quantities $\chi_\rho$ and $\chi_m$ using the HO spectrum.}
    \label{fig:2dplots-ho}
\end{figure}

\begin{figure}[thb!]
    \centerline{
    \includegraphics[width=.9\columnwidth]{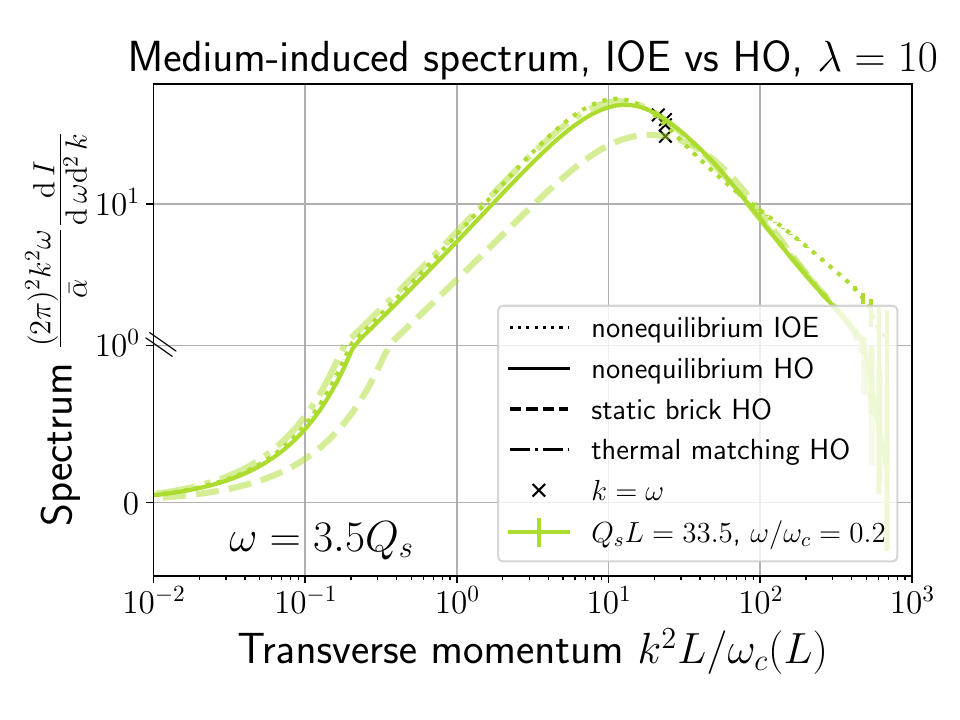}}
        \centerline{
    \includegraphics[width=.9\columnwidth]{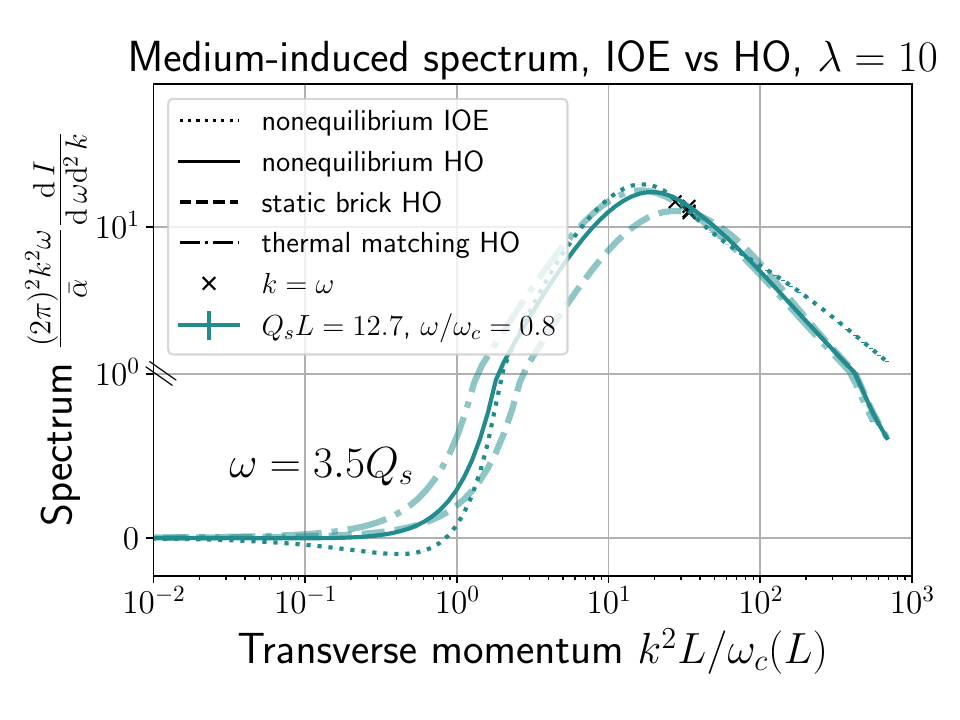}}
        \centerline{
    \includegraphics[width=.9\columnwidth]{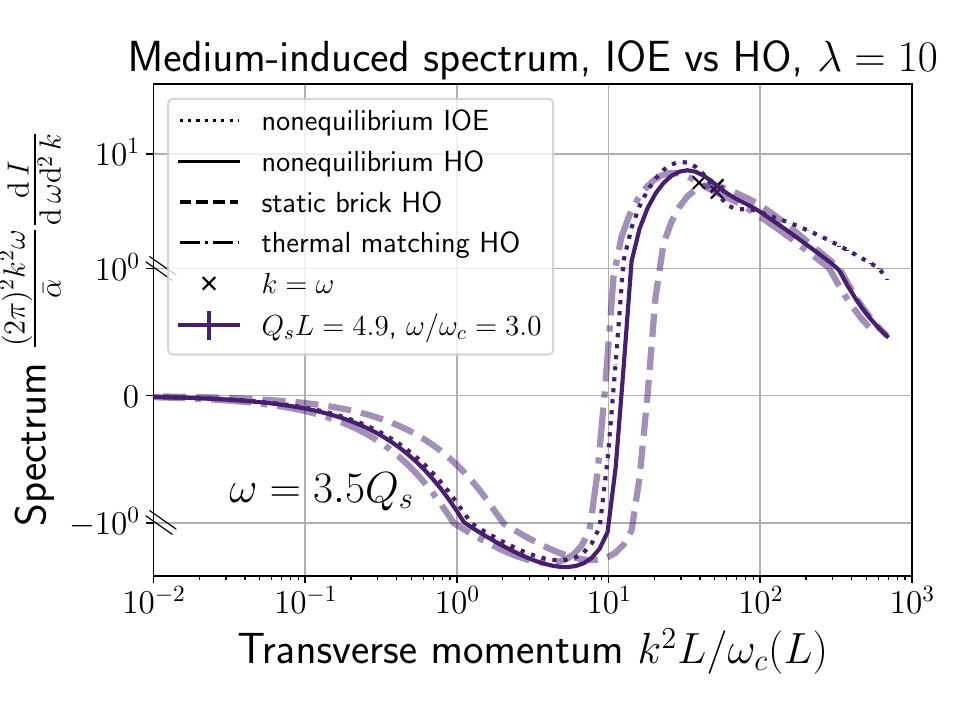}
    }
    \caption{Differences in the spectra obtained using the IOE and HO approximations. Shown are results for $\omega=3.5Q_s$ and varying lengths, displayed in different panels. The lengths correspond to those shown in the bottom-left panel of Fig.~\ref{fig:bjorken}.}
    \label{fig:ioe-ho-spectracomparisons}
\end{figure}

We now move on to the study of Bjorken expanding plasmas, using as input the $\qhat$ values from Ref.~\cite{Boguslavski:2023alu}, as described in Section~\ref{sec:theory}.
In contrast to the static case discussed above, this setup provides a more suitable model description for the initial nonequilibrium stages in HICs.

The evolution of the bulk matter during this stage can be qualitatively divided into several different stages. The system starts from an initially over-occupied state, corresponding to the region before the star marker in Fig.~\ref{fig:qhat_plots}, which is modeled after the distribution function obtained in Glasma simulations \cite{Lappi:2011ju, Kurkela:2015qoa}. It then quickly dilutes and becomes under-occupied while remaining highly anisotropic. In this regime, between the star and triangle markers in Fig.~\ref{fig:qhat_plots}, the nonequilibrium value of $\hat q_0$ is smaller than in equilibrium. The screening mass $\mu_\ast$ likewise reaches its minimum with respect to its thermal evolution in this region. Beyond the circle marker, the system isotropizes, and when approaching the triangle marker, where the pressure anisotropy drops below $2$, it becomes nearly isotropic. At this stage, the nonequilibrium input values of $\hat q_0$ and $\mu_\ast$ approach their thermal limits.

As in the isotropic case, we consider a plasma of extent $L$, which yields the effective time evolution, 
\begin{align}
	\qhat(\tau)&=\begin{cases}
		\qhat^{\mathrm{sim}}(\tau), &  \tmin \leq \tau \leq \tmax\\
		0, & \tau > \tmax\,,
	\end{cases}\label{eq:effective-qhat-evolution-expansion}
\end{align}
with $\tmax=L+\tmin$ as before.
The initial time is fixed to $\tmin=1/Q_s$, and we vary the final time, considering media of different length $L$. Figure~\ref{fig:bjorken} shows the resulting spectra at fixed $\omega$, together with the jet-shape ratio $\chi_\rho$ and the ratio of spectrum maxima $\chi_m$, for three different couplings, $\lambda\in\{0.5,2,10\}$. 

The nonequilibrium spectra qualitatively follow their thermal and static brick counterparts, exhibiting more pronounced deviations at smaller couplings and for larger media, see Fig.~\ref{fig:bjorken}.
The ratio observables $\chi_i$ in Fig.~\ref{fig:bjorken} further demonstrate the interplay of initially over- and later underoccupancy. For small medium lengths $L$, the ratio observable is above unity, reminiscent of the isotropic overoccupied toy models depicted in the lower row of Fig.~\ref{fig:results-iso}. For increasing medium lengths, we observe $\chi_i<1$, similar to the underoccupied toy model.
Strikingly, in contrast to the isotropic (non-expanding) case, the ratios $\chi_i$ do not approach $1$ as $L\to\infty$. This behavior is well expected for a jet propagating in an expanding matter, since $\qhat\to0$ for $L\to\infty$. Indeed, in both nonequilibrium and equilibrium setups, the jet interacts at earlier times with a denser background characterized by a larger $\hat q$, leading to stronger modifications, while its later evolution proceeds in increasingly diluted matter and affects the jet structure less. Consequently, in an expanding background, the jet carries imprints of the medium's initial evolution, which differs between the nonequilibrium and equilibrium cases. This indicates that jets can serve as sensitive tomographic probes of fine features of the medium's evolution, including its thermalization.

\subsection{IOE comparison to HO approximation\label{sec:ioe-ho-comparison}}
In this section, we compare the results from the IOE at the first subleading order as discussed so far with those obtained at leading order. 
This HO limit corresponds to the leading-order term ($\delta v=0$) in the IOE framework, with the scales $Q$ fixed to specific values $Q_b$ and $Q_r$
~\cite{Barata:2020sav}. 
We compare our results for the expanding case of the bottom-up thermalization picture and choose $\lambda=10$ as a characteristic value of the coupling.

We first show the ratios of jet shapes and spectrum maxima in Fig.~\ref{fig:2dplots-ho} for the HO spectrum, as compared to their thermal counterparts, using an illustrative  choice of parameters. 
One should notice the large degree of similarity of these ratios to the corresponding full IOE results shown in the bottom row of Fig.~\ref{fig:bjorken}. 
Direct comparisons of the spectra obtained in the HO approximation and within the IOE framework are presented in Fig.~\ref{fig:ioe-ho-spectracomparisons}, where we show the spectrum as a function of transverse momentum $k^2$ for fixed $\omega$ but different lengths $L$. The static-brick and thermal-matched solutions corresponding to the HO case are also included for reference. We find that for small $k$, the IOE and HO curves are in good agreement, while at larger $k$ they exhibit distinct power law scalings. This difference follows directly from the Coulomb tail in Eq.~\eqref{eq:potential_expansion} accounted for in the IOE framework, and our results are in qualitative agreement with those reported in~\cite{Barata:2021wuf} for a static background.

\subsection{Understanding the influence of the initial stages}
\begin{figure}
    \centering
    \includegraphics[width=\linewidth]{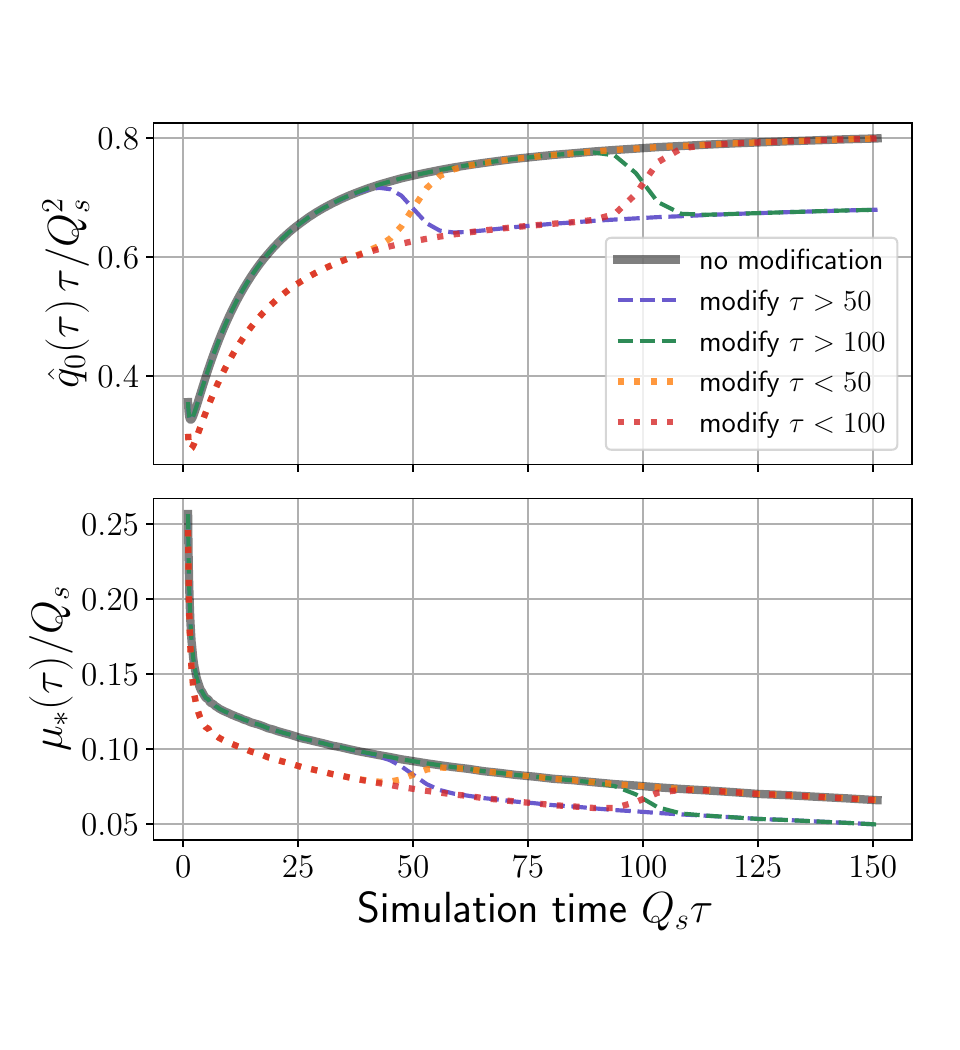}
    \includegraphics[width=\linewidth]{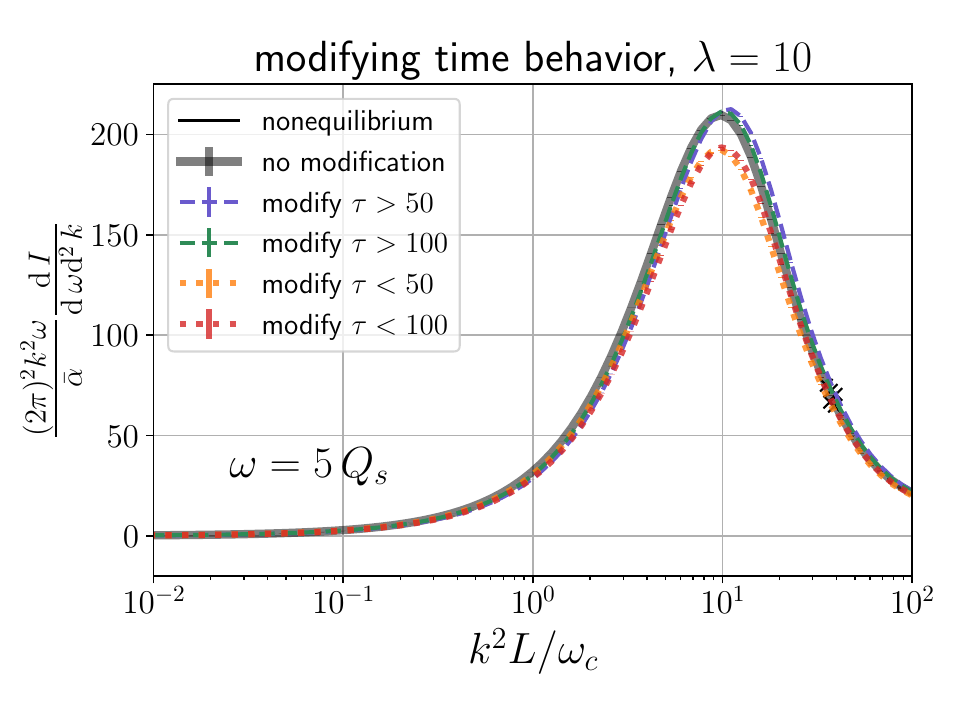}
    \caption{Time evolution of $\hat q_0$ and $\mu_\ast$ (top), with slight modifications applied before or after certain times (see legend), and the resulting spectrum (bottom).
    }
    \label{fig:modifying-time-behavior}
\end{figure}

To conclude this section, we quantify the influence of the initial stages on the overall spectrum, which combines the elements discussed above. Specifically, we use the nonequilibrium evolution of $\hat q_0(\tau)$ and $\mu_\ast(\tau)$ from the expanding EKT simulations at coupling $\lambda=10$, and additionally modify their early- or late-time behavior to examine the impact on the induced spectrum. The top two panels of Fig.~\ref{fig:modifying-time-behavior} show the corresponding input parameters $\hat q_0(\tau)$ and $\mu_\ast(\tau)$. The unmodified evolution is shown by a thick gray line, while early-time modifications are represented by dotted orange and red lines, and the late-time modifications by dashed blue and green lines.

The resulting spectra are displayed as functions of the transverse momentum in the bottom panel. These modifications can lead to significant changes, especially near the peak of the spectrum. Crucially, only the early-time variations have a noticeable effect; modifications applied at later times, $\tau\gtrsim 50/Q_s$, leave the spectrum essentially unchanged. Two clearly distinguishable classes of spectra emerge---those with (dotted lines) and without early-time modifications---demonstrating that the early-time evolution is the dominant factor shaping the spectrum.

Physically, this can be understood from the fact that variations in $\hat q$ are most relevant when $\hat q$ itself is large; at later times, when $\hat q$ becomes smaller, quenching is weak and, thus, further changes have little effect. This naturally explains why the ratio observables $\chi_i$ in Fig.~\ref{fig:bjorken} do not approach unity for expanding systems: although their late-time behavior is identical, they remain sensitive to early-time differences.

\section{Conclusions}
\label{sec:conclusion}
We have conducted the first quantitative study of jet substructure modifications in the presence of out-of-equilibrium QCD matter. To this end, we combined the IOE framework with a realistic nonequilibrium medium evolution obtained from QCD kinetic theory. In particular, we computed the time-dependent transport parameter $\hat q(\Lambda_\perp, \tau)$ in its large-cutoff form in kinetic theory and related it to the bare jet quenching parameter $\hat q_0(\tau)$ and screening mass $\mu_*(\tau)$. Using these, we calculated the differential gluon radiation spectrum at the first sub-leading order in the IOE numerically. 
This approach enabled a self-consistent description of medium-induced gluon radiation in systems that evolve dynamically toward equilibrium.

We considered three cases for the kinetic evolution: an initially over-occupied 
or under-occupied isotropic non-expanding gluonic plasma, and an initially over-occupied longitudinally expanding gluonic plasma undergoing the bottom-up hydrodynamization process in heavy-ion collisions. For each case, we computed the radiation spectrum for the respective nonequilibrium system, for the corresponding (Landau-matched) thermal system with the same energy density, and for a matched static brick, as often used in phenomenological studies of gluon radiation. We observe similar differential spectra, with the static brick capturing the full evolution results less accurately. To quantify differences between the resulting nonequilibrium and thermal radiation spectra, we focused on ratios $\chi_i$ of characteristic quantities like maxima of these spectra or jet shapes as functions of the traversed path, which can be taken as a proxy for time $\tau$. 

Our results demonstrate that, in contrast to isotropic systems, the nonequilibrium stages of the expanding medium leave a sizable imprint on the gluon radiation spectrum. In particular, modifications of $\hat q_0(\tau)$ and $\mu_\ast(\tau)$ at early times significantly alter the emission pattern, while late-time variations have a subleading impact. Consequently, jets produced during the pre-equilibrium stage retain a memory of the medium's early evolution, as reflected in the persistent deviations of the ratio observables $\chi_i$ from unity. The same conclusion can be drawn in the HO limit, where one truncates the IOE at the leading order. These findings indicate that hard probes can serve as sensitive tomographic tools of the early-time dynamics in HICs. 

This work provides a theoretical basis for incorporating pre-equilibrium dynamics into phenomenological studies of jet suppression and substructure. As a future direction, it would be valuable to incorporate these results into existing computations of jet substructure observables, to better gauge how the current theoretical observations transport to phenomenology. Such potential results could also be applied to motivate or test the possibilities of a jet substructure program in future light-ion runs at the LHC. Finally, the current discussion is completely oblivious to anisotropic structures in the bulk, which might play an important role in jet substructure. 
Taking into account such effects in the current framework will require conceptually new ideas, and we leave this for future studies.

\begin{acknowledgements}
    This work is funded in part by the Austrian Science Fund (FWF) under Grant DOI 10.55776/P34455 (KB, FL), and Grant DOI 10.55776/J4902 (FL). FL was a recipient of a DOC Fellowship of the Austrian Academy of Sciences at TU Wien (project 27203). The work of AVS is supported by Fundação para a Ciência e a Tecnologia (FCT) under contract 2022.06565.CEECIND and by the Basque Government through grant IT1628-22. AVS would also like to acknowledge support from Ikerbasque, Basque Foundation for Science. This work has been supported by STRONG-2020 ``The strong interaction at the frontier of knowledge: fundamental research and applications'' which received funding from the European Union’s Horizon 2020 research and innovation programme under grant agreement No 824093. We thank ECT* for support at the Workshop ``New jet quenching tools to explore equilibrium and non-equilibrium dynamics in heavy-ion collisions'' during which this work was initiated. The results in this paper have been achieved in part using the Austrian Scientific Computing (ASC) infrastructure, project 71444. For the purpose of open access, the authors have applied a CC BY public copyright license to any Author Accepted Manuscript (AAM) version arising from this submission.
\end{acknowledgements}
\appendix

\section{Details on the numerical evaluation for the spectrum}
In this appendix, we provide more details on the numeric implementation of Eqs.~\eqref{eq:lo-spectrum} and \eqref{eq:nlo-spectrum} from the main text, which is publicly available \cite{lindenbauer_2025}.

For a given time evolution of $\hat q_0(\tau)$ and $\mu_\ast(\tau)$, we first obtain the matching scales $Q_b$ and $Q_r(\tau)$. For $Q_b$, we rewrite Eq.~\eqref{eq:Qb_final} to find the roots of the function
\begin{align}
    f(Q_b)&=\frac{Q_b^2}{2}
    +
    \int_{\tmin}^{\tmax}\dd{t}\, \hat q_0(t)\log \frac{\mu_\ast(t)}{Q_b}\, .
\end{align}
The minimum of this function is at $Q_b=\sqrt{\int_{\tmin}^{\tmax}\dd{t} \,\hat q_0(t)}$. There only exist solutions for $Q_b$ if the function evaluated at the minimum is negative. We then determine the largest root by using this minimum as the lower bound in a bisection search. Similarly, we find $Q_r$ by searching for the largest root of
\begin{align}
    f(Q_r)=Q_r^4-\hat q_0(t)\omega\log (Q_r^2/\mu_\ast^2) \, ,
\end{align}
which has its extremum at $Q_r=(\hat q_0(t)\omega/2)^{1/4}$.

Next, we construct $\hat q_r(t)=\hat q_0(t)\log\frac{Q_r^2(t)}{\mu_\ast^2(t)}$ to obtain $
    \Omega^2(t)=-\frac{i}{2\omega}\hat q_r(t)$,
and implement a spline interpolation from the known grid points.
We then solve the differential equation~\eqref{eq:differential-equation} for $S$ and $C$ using $\Omega^2$,
\begin{equation}
\label{eq:cs-ho-equations}
\begin{split}
&\left[\frac{\rmd^2}{\rmd t^2}+\Omega^2(t)\right]\{S,C\}(t,t_0)=0  \, , 
\end{split}
\end{equation}
with the boundary conditions $S(t_0,t_0)= \partial_t C(t,t_0)_{t=t_0} =0$ and $C(t_0,t_0)=\partial_t  S(t,t_0)_{t=t_0}=1$.

To estimate the numeric uncertainty of our method, for every spectrum we use four sets of time discretizations, for which we perform the time integrals. The extrapolation to infinite grid points is discussed later.
For every time discretization, we evaluate the functions $S(t_1,t_2)$ and $C(t_1,t_2)$ using
\begin{align}\label{eq:linear_rel_C_S}
S(t_2,t_1)&=C(t_1,t_0)S(t_2,t_0)-S(t_1,t_0)C(t_2,t_0) \, ,\\ 
C(t_2,t_1)&=-\partial_{t_1}C(t_1,t_0)S(t_2,t_0)+\partial_{t_1}S(t_1,t_0)C(t_2,t_0)\notag \, ,
\end{align}
and use these functions evaluated at the specific grid to obtain all other functions, calculating the integral using Simpson's rule.
We additionally employ the analytic limits
\begin{align}
    \lim_{b_r\to\infty}I_a(b_r+i b_i, y)&=-8, \nn
    \lim_{b_r\to\infty}I_b(b_r+i b_i,y)&=4, 
\end{align}
for $b_i,b_r\in\mathbb R,\, y\in\mathbb C$.

The continuum extrapolation to zero grid spacing is performed using that the true value of the observable $\mathcal O$ differs from its evaluation using a fixed grid $\Delta x_i$ by
\begin{align}
    \mathcal O=\mathcal O(\Delta x_1)+a\Delta x_1^n=\mathcal O(\Delta x_2)+a\Delta x_2^n \, ,
\end{align}
with an exponent $n$ depending on the order of the method.
With that, given two spacings $\Delta x_i$, we obtain
\begin{align}
    \mathcal O=\mathcal O(\Delta x_1)+\frac{\mathcal O(\Delta x_2)-\mathcal O(\Delta x_1)}{\Delta x_1^n-\Delta x_2^n}\Delta x_1^n.
\end{align}
For Simpson's rule, one has that $n=4$, except for the double integral (first line of Eq.~\eqref{eq:nlo-spectrum}), for which we obtain only $n=1$ due to our numerical treatment of the double integration. Nevertheless, we have checked that the extrapolation using the four different sets of $\Delta t_i$ lead to nearly identical results. Their differences are indicated by the error bars in the spectra, which are negligible for most parts of the parameter space, indicating the good convergence of the extrapolation.

\bibliographystyle{bibstyle}
\bibliography{references.bib}

\end{document}